\pdfoutput=1

\documentclass[]{iopart}
\usepackage{iopams, graphicx}
\usepackage[pdftex]{hyperref}

\newcommand{\half}{{\textstyle \frac{1}{2}}}

\newcommand{\C}{\mathcal{C}}
\newcommand{\U}{\mathcal{U}}
\newcommand{\F}{\mathcal{F}}

\newcommand{\plot}[1]{     
  \begin{minipage}[t]{0.5\textwidth}
    \includegraphics[width=\textwidth]{#1}
  \end{minipage}
}

\begin{document}

\title{Testing outer boundary treatments for the Einstein equations}
\author{Oliver Rinne, Lee Lindblom and Mark A. Scheel}
\address{Theoretical Astrophysics 130-33, California Institute of Technology, 
         Pasadena, CA 91125, USA}

\begin{abstract}
  Various methods of treating outer boundaries in numerical relativity
  are compared using a simple test problem: a Schwarzschild black hole
  with an outgoing gravitational wave perturbation.  Numerical
  solutions computed using different boundary treatments are compared
  to a `reference' numerical solution obtained by placing the outer
  boundary at a very large radius.  For each boundary treatment, the
  full solutions including constraint violations and extracted
  gravitational waves are compared to those of the reference solution,
  thereby assessing the reflections caused by the artificial boundary.
  These tests are based on a first-order generalized harmonic formulation of
  the Einstein equations and are implemented using a pseudo-spectral
  collocation method.
  Constraint-preserving boundary conditions
  for this system are reviewed, and an improved boundary condition on
  the gauge degrees of freedom is presented.  Alternate boundary
  conditions evaluated here include freezing the incoming
  characteristic fields, Sommerfeld boundary conditions, and the
  constraint-preserving boundary conditions of Kreiss and Winicour.
  Rather different approaches to boundary treatments, such as sponge
  layers and spatial compactification, are also tested.  Overall the
  best treatment found here combines boundary conditions that preserve the
  constraints,  
  freeze the Newman-Penrose scalar $\Psi_0$, and control gauge reflections.
\end{abstract}



\section{Introduction}
\label{s:Introduction}

A fundamental problem in numerical relativity is the
need to solve Einstein's equations on spatially
unbounded domains with finite computer resources. 
There are various ways of addressing this
issue. Most often, the spatial domain is truncated at a finite
distance and suitable boundary conditions are imposed at the
artificial boundary. A different approach is to compactify the domain by using
spatial coordinates that bring spatial infinity to a
finite location on the computational grid. Another method often used 
for wave-like problems (although it is not
commonly used in numerical relativity) includes so-called
\emph{sponge layers}  which damp the waves
near the outer boundary of the computational domain.  The purpose 
of this paper is to compare these various methods by testing 
their ability to accurately reproduce dynamical solutions of 
Einstein's equations.

An ideal boundary treatment would produce a solution to Einstein's
equations that is identical (within the computational domain) to the
corresponding solution obtained on an unbounded domain. 
In particular, no spurious gravitational radiation or constraint
violations should enter the computational domain through the
artificial boundary. We can use this principle to test the various
boundary treatments in the following way. 
First we compute a \emph{reference solution} using a very large 
computational domain, large enough that its boundary remains out of 
causal contact with the interior spacetime region where
comparisons are being made.
Next we compute the same solution using a domain truncated
at a smaller distance where one of the boundary treatments is used:
we either impose boundary conditions there, compactify spatial infinity, or add
a sponge layer.
Finally we compare the solution on the smaller domain with the reference
solution, measuring the reflections and constraint
violations caused by the boundary treatment.
Assessing boundary conditions by comparing with a reference solution
on a much larger domain or a known analytic solution is a common
practice in computational science. For applications to numerical
relativity see e.g. \cite{Novak2004}, chapter 8 of
\cite{RinnePhD}, and \cite{Lau2004a, Babiuc2006a, Babiuc2007}.

The particular test problem used in this paper is a Schwarzschild
black hole with an outgoing gravitational wave perturbation.
The interior of the black hole is excised; all the characteristic
fields propagate into the black hole (and out of the computational domain) 
at the inner boundary and hence no boundary
conditions are needed there.
Our numerical implementation uses a pseudo-spectral collocation method.
See \ref{s:NumericalDetails} for details on the initial data,
the numerical methods, and the quantities that we compare
between the solutions.

We perform all of these tests using a first-order generalized harmonic
formulation of the Einstein equations (see \cite{Lindblom2006} and 
references therein).  In section \ref{s:CPBC} we
discuss the construction of boundary conditions for this system that
prevent the influx of constraint violations, 
and that limit the spurious incoming gravitational radiation 
by controlling the Newman-Penrose scalar $\Psi_0$ at the boundary.
We also improve the boundary conditions on the gauge degrees of
freedom by studying small gauge perturbations of flat spacetime.  We
then evaluate the performance of these boundary
conditions on our test problem: measuring the reflections and
constraint violations caused by
the computational boundary, and determining how these
reflections vary with the radius of the boundary.

Section \ref{s:AltBC} evaluates the performance of a variety of other
widely used boundary conditions on our test problem.
First we test the simple boundary conditions that freeze
all the incoming characteristic fields at the boundary.  We also test
the commonly used variant of this, the Sommerfeld boundary conditions,
used in many binary black hole simulations~\cite{Brugmann2004,
Campanelli2006,Baker2006, Diener2006, Herrmann2006} based on the 
BSSN~\cite{Shibata1995,Baumgarte1998} formulation of Einstein's equations.
Finally in section \ref{s:AltBC} 
we evaluate the constraint-preserving boundary conditions
proposed by Kreiss and Winicour \cite{Kreiss2006}, which differ from
those discussed in section \ref{s:CPBC} mainly by our use of a physical
boundary condition that controls $\Psi_0$.

In section \ref{s:AltApproach} we evaluate two boundary treatments that
are alternatives to imposing local boundary conditions at a finite
outer boundary.  The first is the spatial compactification method used 
e.g.~by Pretorius~\cite{Pretorius2005a,Pretorius2005b,Pretorius2006} in his
groundbreaking binary black hole evolutions.  In this treatment a coordinate
transformation maps spatial infinity to a finite location on the
computational grid.  As waves travel out, they
become increasingly blue-shifted with respect to the compactified
coordinates and ultimately they fail to be resolved.  Hence numerical
dissipation is applied, which damps away these short-wavelength features.
We measure the reflections and the constraint violations generated by
the waves in our test problem as they interact with this boundary
treatment.  Finally in section \ref{s:AltApproach} 
we implement and test a sponge layer method for Einstein's equations.

One of the main objectives of current binary black hole simulations is
the computation of reliable waveforms for gravitational wave data
analysis.  Therefore it is important to evaluate how the various boundary
treatments affect the accuracy of the extracted waveforms.  In
section \ref{s:Radiation}, we compute the Newman-Penrose scalar $\Psi_4$
(which describes the outgoing waves) on an extraction sphere close to
the outer boundary (or compactified region, or sponge layer,
respectively) and compare it with the analogous $\Psi_4$ from the
reference solution.  We also compare the measured reflections caused by our
$\Psi_0$ controlling boundary condition with the analytical
predictions of these reflections made by 
Buchman and Sarbach~\cite{Buchman2006,Buchman2007}.

Finally we discuss the implications of our results in
section \ref{s:Discussion}, and we also describe briefly a number of other 
boundary treatments which we do not test here.


\section{Constraint-preserving boundary conditions}
\label{s:CPBC}

In this section, we briefly review the generalized harmonic form of
the Einstein evolution system used in our tests.  The method of
constructing constraint-preserving boundary conditions (CPBCs) for
this system is also discussed, and an improved boundary condition for
the gauge degrees of freedom is derived.  The numerical performance of
these boundary conditions is evaluated using our test problem, and the
dependence of the spurious reflections as a function of the boundary
radius is measured.

\subsection{The generalized harmonic evolution system}
\label{s:GHSystem}

The formulation of Einstein's equations employed here uses generalized
harmonic gauge conditions, in which the coordinates $x^a$ obey the 
wave equation
\begin{equation}
  \label{e:BoxX}
  \Box x^a = H^a (x, \psi),
\end{equation}
where $\Box=\psi^{ab}(\partial_a\partial_b - \Gamma^c{}_{ab}\partial_c)$
is the covariant scalar wave operator, with $\psi_{ab}$ the spacetime metric
and $\Gamma^c{}_{ab}$ the associated metric connection.
In this formulation of the Einstein system the gauge source function 
$H^a$ may be chosen freely as a function of the coordinates and of the
spacetime metric $\psi_{ab}$ (but not derivatives of $\psi_{ab}$).

As is well known, the Einstein equations reduce to a set of coupled wave
equations when the gauge is specified by equation \eref{e:BoxX}.
We write this system in first-order form, both in time and space, by
introducing the additional variables $\Phi_{iab} \equiv \partial_i \psi_{ab}$ 
and $\Pi_{ab} \equiv -t^c \partial_c \psi_{ab}$, where $t^c$ is the future
directed unit normal to the $t=\mathrm{const.}$ hypersurfaces.
Here lower-case Latin indices from the beginning of the alphabet denote 
four-dimensional spacetime quantities, whereas
lower-case Latin indices from the middle of the alphabet are spatial.
The principal parts of these evolution equations are 
given by \footnote{The parameter $\gamma_1$ of \cite{Lindblom2006} 
  is chosen to be $-1$, which ensures that the equations are linearly 
  degenerate.}
\begin{eqnarray}
  \partial_t \psi_{ab} \simeq 0, \nonumber\\
  \label{e:ghev}
  \partial_t \Pi_{ab} \simeq N^k \partial_k \Pi_{ab} - N g^{ki}
      \partial_k \Phi_{iab} - \gamma_2 N^k \partial_k \psi_{ab}, \\ 
  \partial_t \Phi_{iab} \simeq N^k \partial_k \Phi_{iab} - N
      \partial_i \Pi_{ab} + N \gamma_2 \partial_i \psi_{ab},\nonumber
\end{eqnarray}
where $\simeq$ indicates that purely algebraic terms have been omitted, 
$g_{ij}$ is the spatial metric of the $t = \mathrm{const.}$ slices,
and $N$ and $N^i$ are the lapse function and shift vector, respectively. 
The parameter $\gamma_2$ was introduced in 
\cite{Lindblom2006} in order to damp 
violations of the three-index constraint 
\begin{equation}
  \label{e:C3}
  \C_{iab} \equiv \partial_i \psi_{ab} - \Phi_{iab} = 0. 
\end{equation}
We also include terms of lower derivative order that are designed 
to damp violations of the harmonic gauge constraint \cite{Gundlach2005}
\begin{equation}
  \label{e:C1}
  \C_a \equiv -\Box x_a + H_a = \psi^{bc} \Gamma_{abc} + H_a = 0.
\end{equation}

The system \eref{e:ghev} is symmetric hyperbolic.
The characteristic fields in the direction $n_i$ (where $n_a t^a=0$) 
are given by
\begin{eqnarray}
  u^0_{ab} = \psi_{ab}, & \mathrm{speed}\,\quad 0, \\
  \label{e:u1}
  u^{1\pm}_{ab} = \Pi_{ab} \pm \Phi_{nab} - \gamma_2 \psi_{ab}, \quad
    & \mathrm{speed} \, \quad-N^n \pm N,\\
  \label{e:u2}
  u^2_{Aab} = \Phi_{Aab}, & \mathrm{speed} \, \quad-N^n. 
\end{eqnarray}
For future reference, we also define  
\begin{equation}
  \label{e:u1tilde}
  \tilde u^{1\pm}_{ab} \equiv \Pi_{ab} \pm \Phi_{nab}.
\end{equation}
Here and in the following, an index $n$ denotes contraction with $n_i$,
while upper-case Latin indices $A,B,\ldots$ are orthogonal to $n$,
e.g. $v_A = P_{Ai} v^i$ where $P_{ab} \equiv \psi_{ab} - n_a n_b+t_at_b$.
For further details, we refer the interested reader to
\cite{Lindblom2006}.

\subsection{Construction of boundary conditions}
\label{s:BCConstr}

Our construction of boundary conditions for the generalized harmonic
evolution system can be divided into three parts: constraint-preserving, 
physical, and gauge boundary conditions.

In order to impose constraint-preserving boundary conditions,
we derive the subsidiary evolution system that the constraints
\eref{e:C3} and \eref{e:C1} obey as a consequence of the main evolution
equations \eref{e:ghev}. The incoming modes of the subsidiary system
are then required to vanish at the boundary (cf.~\cite{Stewart1998,
Friedrich1999,Iriondo2002,Calabrese2002,Calabrese2003,Calabrese2003a,
Kidder2005,Bona2005,Sarbach2005}).
For instance, the harmonic gauge constraint \eref{e:C1} obeys a wave 
equation
\begin{equation}
  \label{e:ConstraintEvolution}
  \Box \C_a = \textrm{(lower-order terms homogeneous in the constraints})
\end{equation}
and the corresponding incoming fields will involve first
derivatives of $\C_a$.
In terms of the incoming modes $u^{1-}_{ab}$ \eref{e:u1} of the main 
evolution equations, the resulting constraint-preserving
boundary conditions can be written in the form
\begin{eqnarray}
  \label{e:CPBC}
  P_{ab}^{\mathrm{C}\,cd} \partial_n u^{1-}_{cd} 
  &\equiv& (\half P_{ab} P^{cd} - 2 l_{(a} P_{b)}{}^{(c}k^{d)} + l_a l_b
    k^c k^d) \partial_n u^{1-}_{cd} \nonumber\\
  &\doteq& (\textrm{tangential derivatives}),
\end{eqnarray}
where $P^\mathrm{C}$ is a projection operator of rank $4$ 
(cf.~\cite{Lindblom2006}).
Here $n_i$ now refers to the outward-pointing unit spatial normal to the
boundary, $l^a = (t^a + n^a)/\sqrt{2}$, $k^a = (t^a - n^a)/\sqrt{2}$,
and $\doteq$ denotes equality at the boundary.
If the shift vector points towards the exterior at the boundary 
($N^n \, \dot > \, 0$),
the fields $u^2_{Aab}$ \eref{e:u2} are incoming as well and we obtain
a boundary condition on them by requiring the components $\C_{nAab}$
of the four-index constraint
\begin{equation}
  \label{e:C4}
  \C_{ijab} \equiv -2 \partial_{[i} \Phi_{j]ab}
\end{equation}
to vanish at the boundary.

An acceptable physical boundary condition should require that no
gravitational radiation enter the computational domain from the
outside (except for backscatter off the spacetime curvature, an effect
that is a first-order correction in $M/R$).
Gravitational radiation may be described by the evolution system that
the Weyl tensor obeys by virtue of the Bianchi identities 
(see e.g.~\cite{Kidder2005}). 
Our boundary condition requires the incoming characteristic fields of
this system to vanish at the outer boundary.
These incoming fields are proportional to the Newman-Penrose 
scalar $\Psi_0$ 
(evaluated for a Newman-Penrose null tetrad containing the vectors 
$l^a$ and $k^a$). 
Hence the physical boundary condition we use is 
\cite{Kidder2005,Friedrich1999,Bardeen2002,Sarbach2005,Nagy2006}
\begin{equation}
  \label{e:ZeroPsi0}
  \partial_t\Psi_0 \doteq 0,
\end{equation}
which can be written in a form similar to \eref{e:CPBC},
\begin{eqnarray}
  \label{e:PhysBC}
  P_{ab}^{\mathrm{P}\,cd} \partial_n u^{1-}_{cd} 
  &\equiv& (P_a{}^c P_b{}^d - \half P_{ab} P^{cd}) \partial_n u^{1-}_{cd} 
    \nonumber\\
  &\doteq& (\textrm{tangential derivatives}).  
\end{eqnarray}
Here $P^\mathrm{P}$ is a projection operator of rank $2$ that is
orthogonal to $P^\mathrm{C}$ \cite{Lindblom2006}.
We remark that \eref{e:ZeroPsi0} still causes some, albeit very small,
spurious reflections of gravitational radiation. 
It can be viewed as the lowest level in
a hierarchy of perfectly absorbing boundary conditions for linearized
gravity \cite{Buchman2006,Buchman2007}.

The constraint-preserving \eref{e:CPBC} and physical \eref{e:PhysBC} 
boundary conditions together constrain six components of the main incoming
fields $u^{1-}_{ab}$. The remaining four components correspond to gauge
degrees of freedom. In the past we chose simply to freeze those components
in time~\cite{Lindblom2006},
\begin{equation}
  \label{e:OldGaugeBC}
  P_{ab}^\mathrm{G\,cd} \partial_t u^{1-}_{cd} \doteq 0,
\end{equation}
where $P^\mathrm{G} \equiv \mathbb{I} - P^\mathrm{C} - P^\mathrm{P}$.

The initial-boundary value problem (IBVP) for the boundary conditions
discussed so far was shown in \cite{Rinne2006} to be 
boundary-stable, which is a (rather strong) necessary condition for
well posedness. These boundary conditions have been successfully used in
long-term stable evolutions of single and binary black hole spacetimes
\cite{Lindblom2006,Scheel2006,Pfeiffer2007}.
In the following subsection, we present 
an improvement to the gauge boundary condition
(\ref{e:OldGaugeBC})
motivated by the evolution of gauge perturbations about flat spacetime.

\subsection{Improved gauge boundary condition}
\label{s:GaugeBC}

Let us assume that near the outer boundary, the spacetime is close to 
Minkowski space in standard coordinates ($H_a = 0$) so that the Einstein 
equations may be linearized about that background.
This assumption is reasonable because for the dominant wavenumber of
the outgoing pulse ($k = 1.6/M$) and the boundary radius we typically
consider ($R = 41.9 M$), we have $kR \gg 1$ and $R \gg M$.
Furthermore, we assume that the outer boundary is a coordinate sphere
of radius $r = R$.

We begin by noting that harmonic gauge does not fix the coordinates 
completely: infinitesimal coordinate transformations
\begin{equation}
  x^a \rightarrow x^a + \xi^a
\end{equation}
are still allowed
provided the displacement vector satisfies the wave equation,
\begin{equation}
  \label{e:XiWave}
  \Box \xi^a = 0.
\end{equation}
Under such a coordinate transformation, the metric changes by
\begin{equation}
  \label{e:PsiPerturbation}
  \delta \psi_{ab} = - 2 \partial_{(a} \xi_{b)}.
\end{equation}
A closer inspection \cite{Rinne2006} of the projection operator 
$P^\mathrm{G}$ in \eref{e:OldGaugeBC} shows that the gauge boundary 
conditions control the components  $l^a \delta \psi_{ab}$ of the 
perturbations, where $l^a \equiv (t^a + n^a)/\sqrt{2}$ is the outgoing null
vector normal to the boundary.
It is interesting to observe that these components vanish in the 
ingoing radiation gauge \cite{Chrzanowski1975}.
However, imposing radiation gauge on the entire spacetime is not possible 
in spacetimes containing strong-field regions, which will always
generate perturbations $l^a \delta \psi_{ab}$ that propagate into 
the far field.
A reasonable condition to require then is that these perturbations
pass through the boundary without causing strong reflections.

Each Cartesian component of the vector $l^a \delta \psi_{ab}$ obeys
the scalar wave equation
\begin{equation}
  \Box \psi = 0 .
\end{equation}
Solutions to this equation can be written in the form 
\begin{equation}
  \psi = \sum_{l=1}^\infty \sum_{m = -l}^l Y_{lm} (\theta, \phi) \psi_l (t, r),
\end{equation}
where the $Y_{lm}$ are the standard spherical harmonics and the $\psi_l$
are linear combinations of outgoing ($+$) and incoming ($-$) solutions
\begin{equation}
  \psi_l^\pm (t, r) = r^{l-1} \left( \frac{\partial}{\partial r} \frac{1}{r}
  \right)^l F_l^\pm (r \mp t),
\end{equation}
$F^\pm_l(x)$ being arbitrary functions.
A boundary condition is needed  on $\psi$ that
eliminates the incoming part of these solutions. 
In \cite{Bayliss1980}, a hierarchy of boundary conditions is constructed
that accomplish this task for all $l \leqslant L$.
This idea was applied to the evolution of the Weyl curvature in
\cite{Buchman2006} in order to construct improved \emph{physical} 
boundary conditions.
For the gauge boundary conditions considered here, we restrict ourselves
to the $L = 0$ member of the hierarchy, which corresponds to the 
Sommerfeld condition \footnote{
  To avoid confusion, we remark that in  \cite{Babiuc2007, Kreiss2006},
  the term `Sommerfeld condition' is used in reference to
  a condition of the form $(\partial_t + \partial_r) u \doteq 0$, 
  i.e.~\emph{without} the extra $r^{-1}$ term due to our polar coordinates.
}
\begin{equation}
  \label{e:NewGaugeBC1}
  (\partial_t + \partial_r + r^{-1}) \psi \doteq 0.
\end{equation}
In contrast, our old gauge boundary condition that froze 
the incoming characteristic field, as in \eref{e:OldGaugeBC}, 
is given by
\begin{equation}
  \label{e:OldGaugeBC1}
  (\partial_t + \partial_r + \gamma_2) \psi \doteq 0,
\end{equation}
where $\gamma_2$ is the constraint damping parameter.

\begin{figure}
  \plot{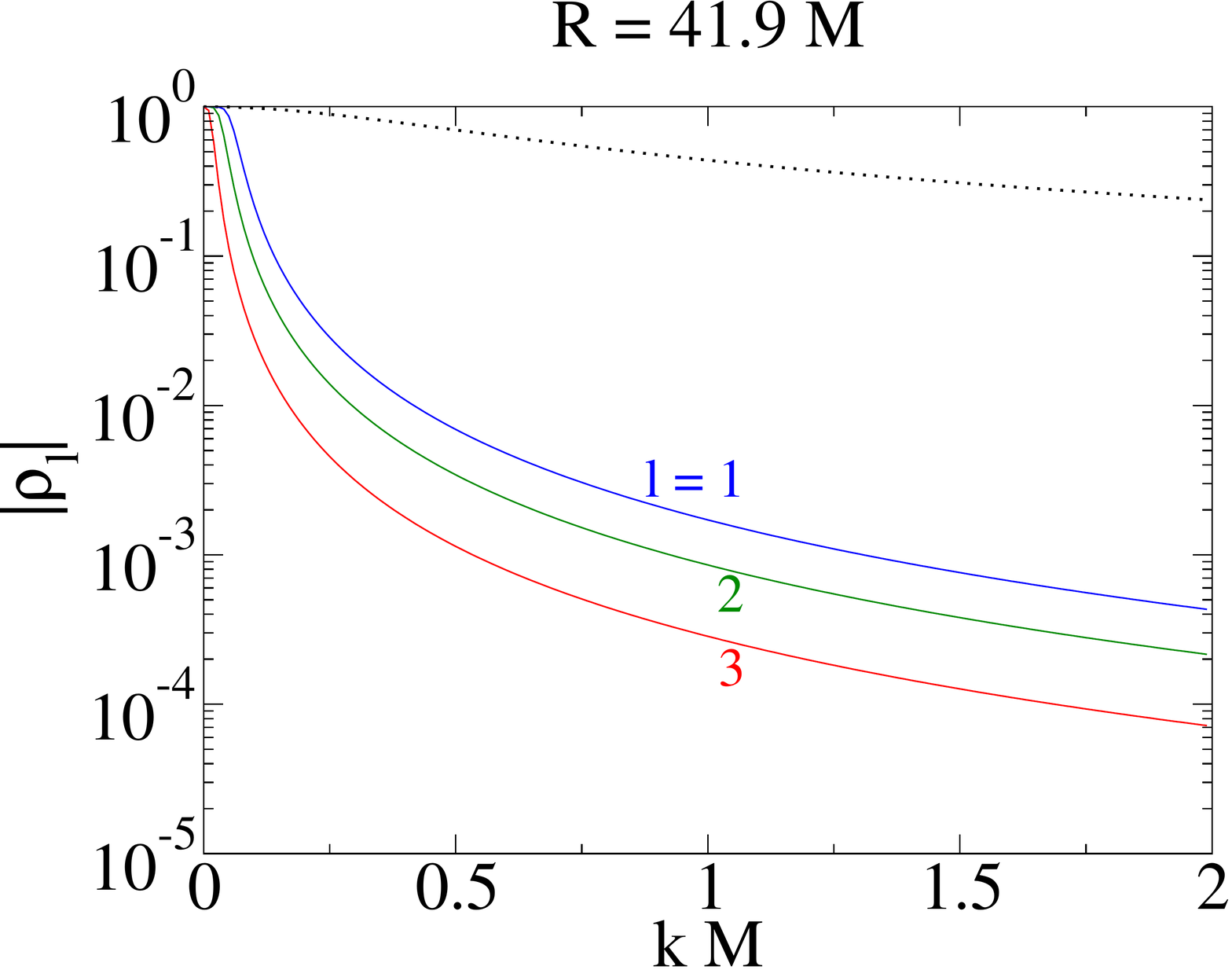}
  \plot{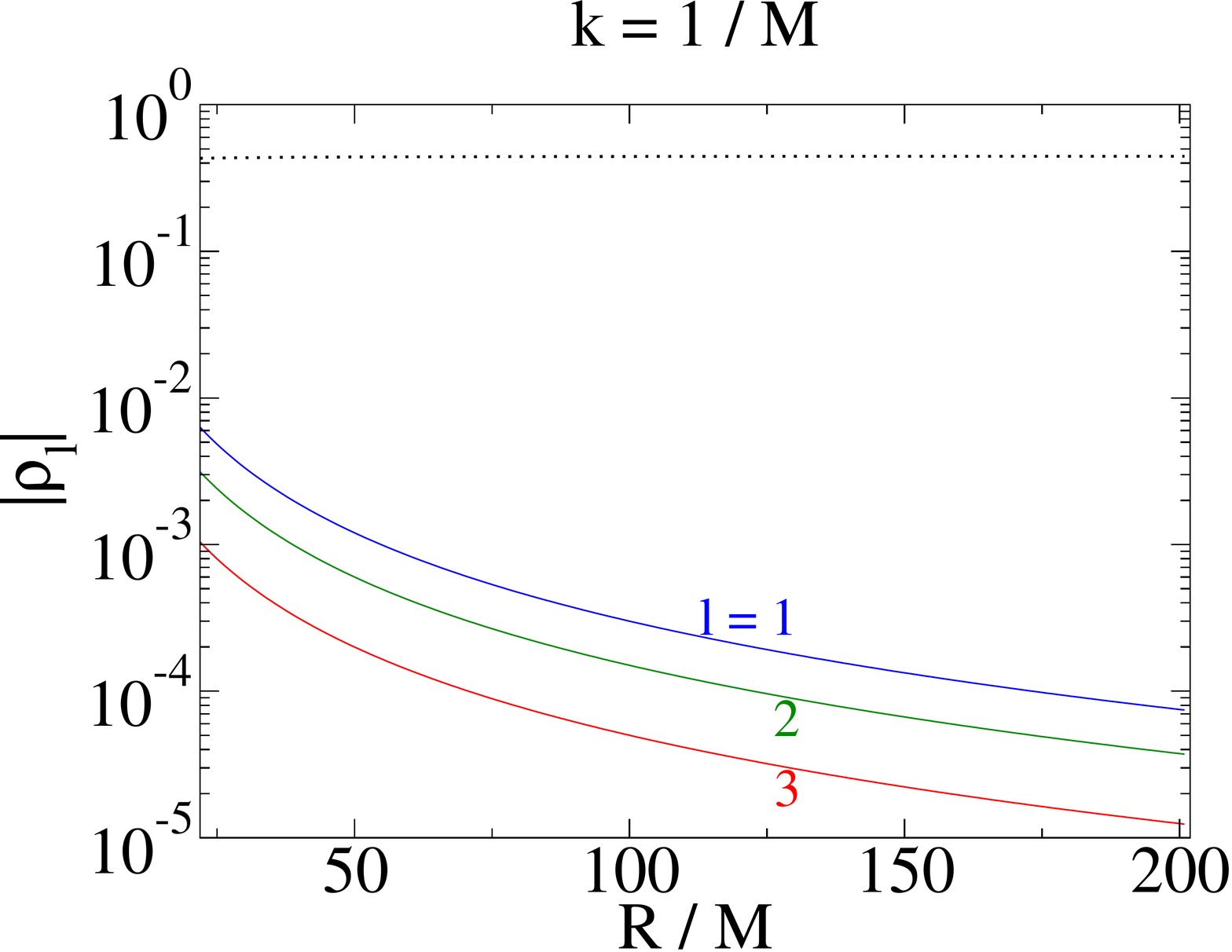}
  \caption{\label{f:ReflCoeff} 
    Predicted reflection
    coefficients $\rho_l$ for freezing (dotted) and Sommerfeld
    (solid) boundary conditions as functions of wave number $k$ and
    outer boundary radius $R$.  The curves for different $l$ are
    visually indistinguishable in the freezing case.
    Note also that $\rho_0 = 0$ for the Sommerfeld condition.
  }
\end{figure}

This Sommerfeld boundary condition \eref{e:NewGaugeBC1} is
much less reflective than the freezing condition \eref{e:OldGaugeBC1}.
To see this, we consider a solution of the form
\begin{equation}
  \psi_l = \psi_l^+ + \rho_l \psi_l^- 
\end{equation}
with generating functions
\begin{equation}
  F^\pm_l (x) = \rme^{\pm ikx} ,
\end{equation}
where $k \in \mathbb{R}$ is the wave number.
Substituting this solution into the boundary conditions
\eref{e:NewGaugeBC1} resp.~\eref{e:OldGaugeBC1}, we solve for the
reflection coefficient $\rho_l$.
Figure \ref{f:ReflCoeff} shows $|\rho_l|$ for a typical range of wave
numbers $k$ and outer boundary radii $R$ used for the numerical tests
in this paper. (The dominant wave number of the outgoing pulse is $k
\approx 1.6/M$ and in most cases, we place the outer boundary at $R = 41.9 M$.)
We see that $|\rho_l|$ is much smaller (by about $3$ orders of
magnitude) for the Sommerfeld condition than for the freezing condition.

In the notation of the previous subsection, the improved gauge
boundary condition \eref{e:NewGaugeBC1} reads (after taking a time derivative),
\begin{equation}
  \label{e:NewGaugeBC}
  P_{ab}^\mathrm{G\,cd} \partial_t 
    [ u^{1-}_{cd} + (\gamma_2 - r^{-1}) \psi_{cd} ] \doteq 0.
\end{equation}
We remark that the extra terms in \eref{e:NewGaugeBC} 
as compared with the old condition \eref{e:OldGaugeBC}
are of lower derivative order, so that the high-frequency stability result of 
\cite{Rinne2006} extends immediately to these modified gauge 
boundary conditions.

\subsection{Numerical results}
\label{s:CPBCComparisons}

\begin{figure}
  \plot{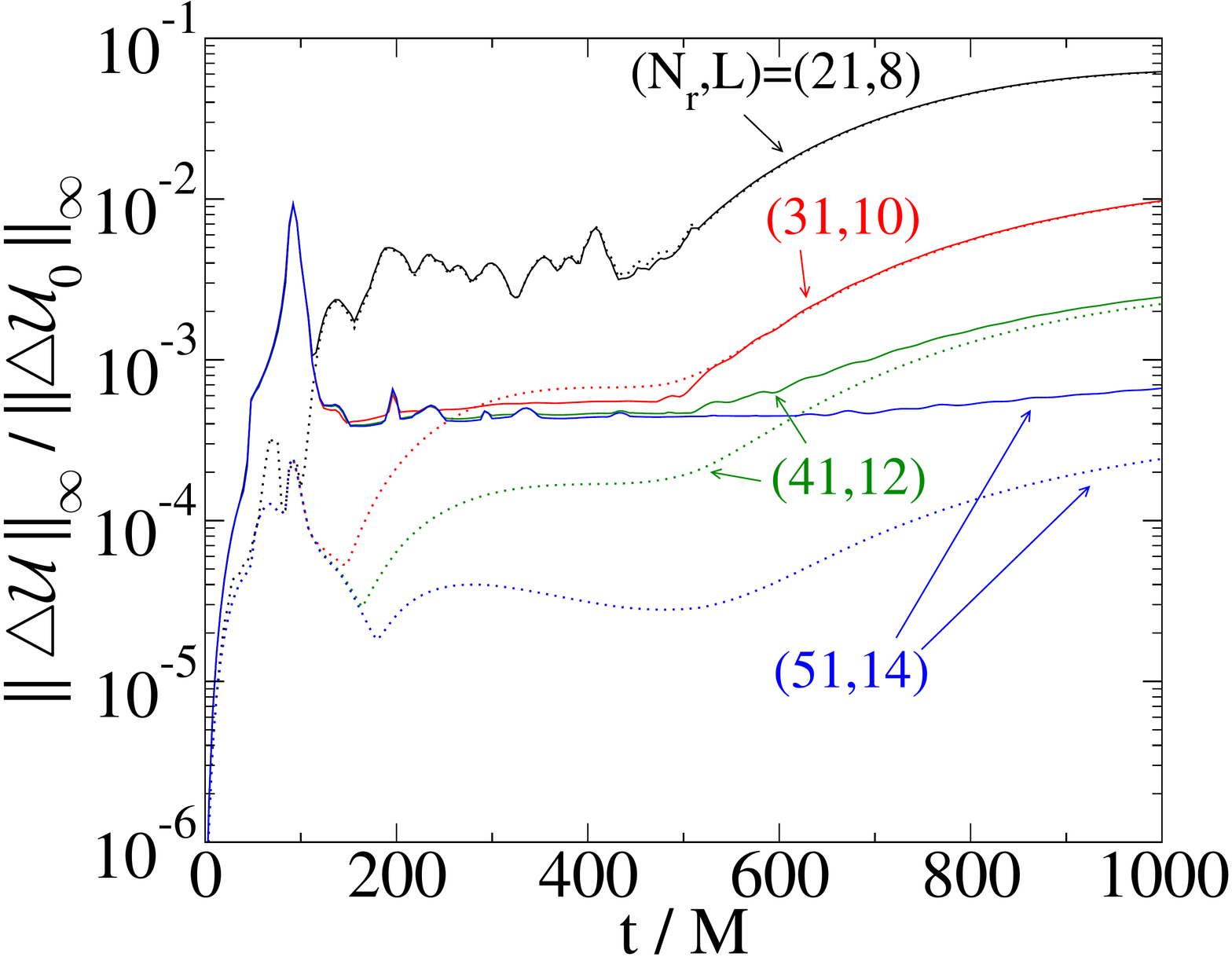}
  \plot{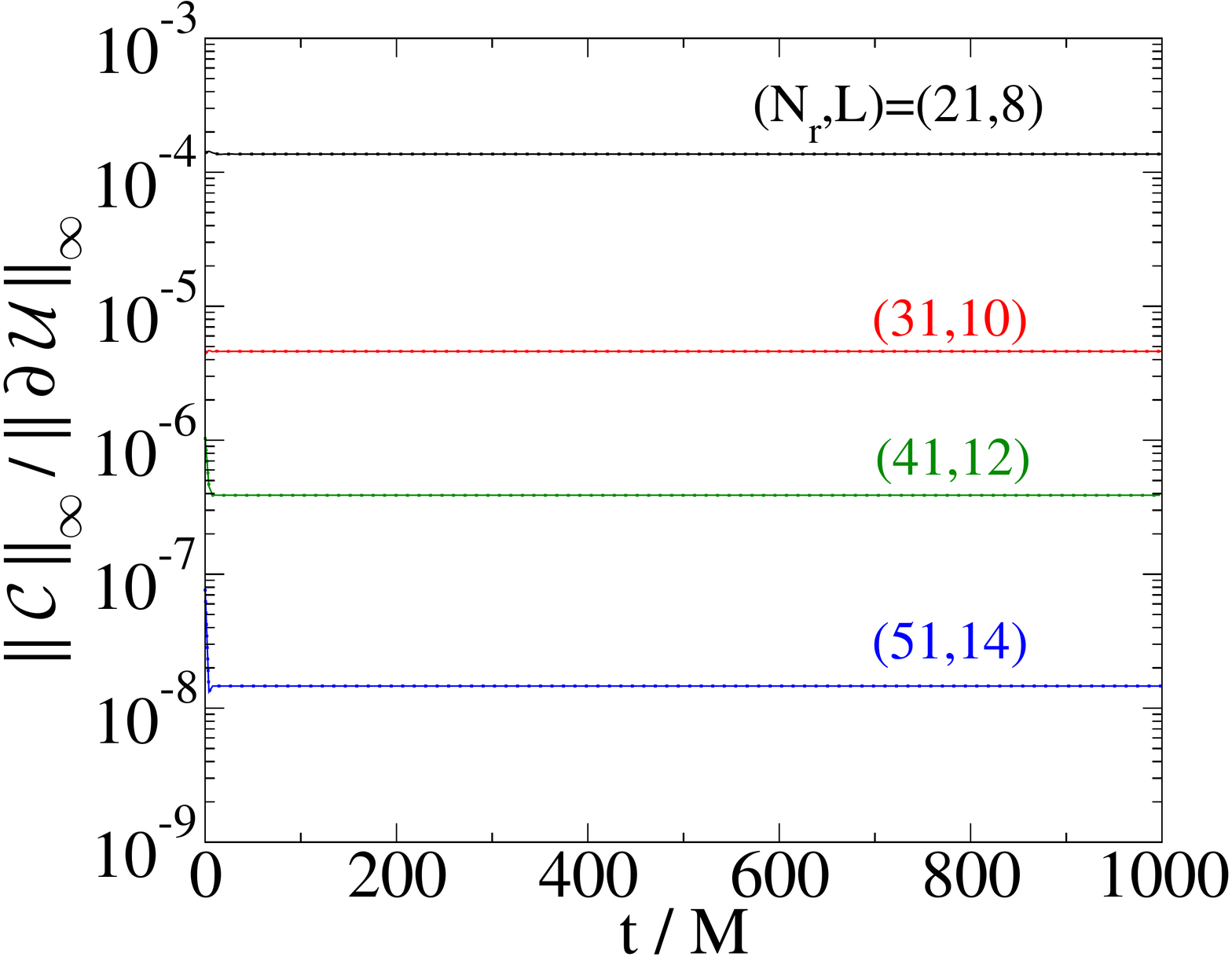}

  \plot{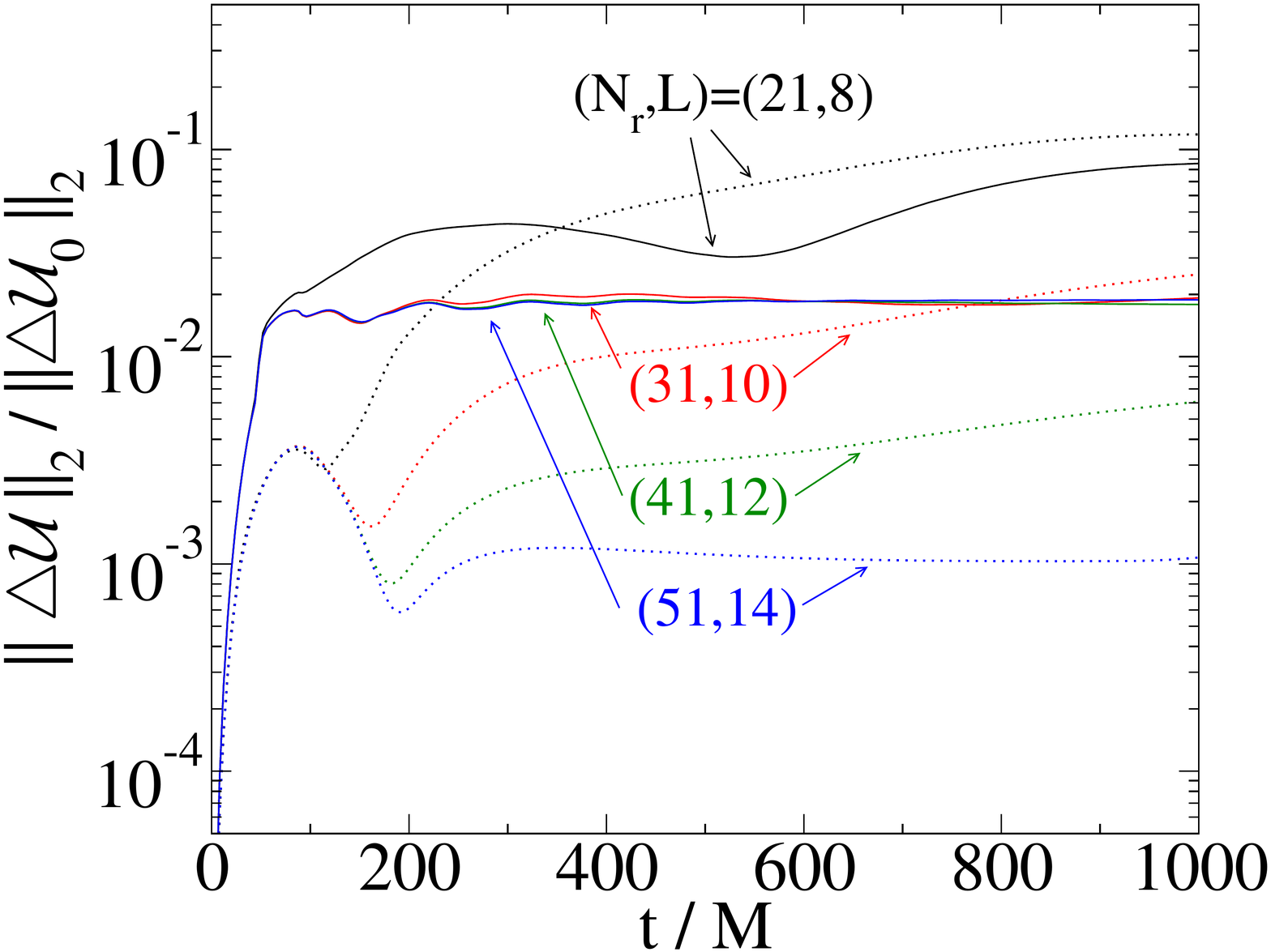}
  \plot{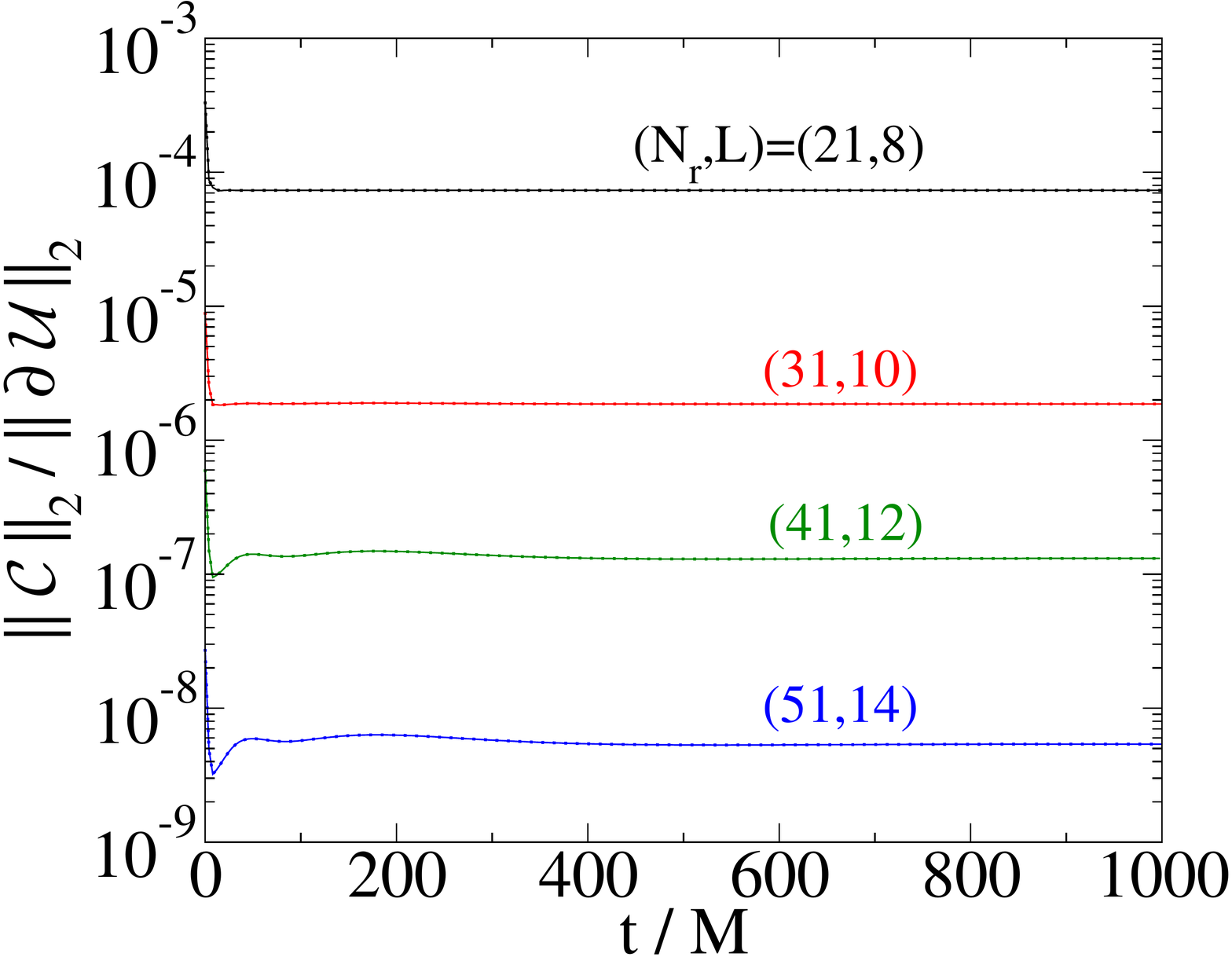}
  \caption{\label{f:OldVsNewCp} 
    Old (solid) vs. new (dotted) CPBCs.
    Four different resolutions are shown: $(N_r, L) = (21, 8)$, 
    $(31, 10)$, $(41, 12)$, and $(51, 14)$.
    The outer boundary is at $R = 41.9 M$.
  }
\end{figure}

The numerical tests of the various boundary conditions performed in this
paper are described in some detail in \ref{s:NumericalDetails}.
Figure \ref{f:OldVsNewCp} compares the numerical performance of our new 
CPBCs \eref{e:CPBC}, \eref{e:C4}, \eref{e:PhysBC}, \eref{e:NewGaugeBC} with 
our old ones \eref{e:CPBC}, \eref{e:C4}, \eref{e:PhysBC}, \eref{e:OldGaugeBC}.
The outer boundary is placed at radius $R = 41.9 M$ for these particular tests.
Shown are the discrete $L^\infty$ and $L^2$ norms of the difference 
$\Delta \U$ between the numerical solution and the reference 
solution, and also the violations of the constraints $\C$
(see \ref{s:ErrorQuantities} for precise definitions of these quantities).
The reference solution has an outer boundary at radius $961.9M$ and is 
computed using our old CPBCs; thus for $t<920M$ the outer boundary of the
reference solution is out of causal contact with the region where
$\Delta \U$ and $\C$ are computed.

In the difference $\Delta \U$ we see a reflection that originates when the 
wave reaches the boundary at $t \approx R$ and then amplifies as it
moves inward in the spherical geometry, 
assuming its maximum at $t \approx 2 R$.
This feature is much more prominent in the $L^\infty$ norm than in the
$L^2$ norm, which is why we display only the $L^\infty$ norm in
subsequent plots.
The reflection is much smaller (by a factor of $\approx R/M$) for the new
boundary conditions as compared with the old ones.
Even at later times, the new boundary conditions result in a smaller
$\Delta \U$, which in contrast to the old conditions appears to
decrease as resolution is increased. 

We would like to stress that $\Delta \U$ is a coordinate dependent quantity.
Hence a smaller $\Delta \U$ does not necessarily mean that the
boundary treatment is `better' in a physically meaningful sense.
If however the aim is to produce a solution that is as close to the
reference solution in the same coordinates, the choice of gauge
boundary conditions does become important.
Gauge reflections can in principle also impair the numerical accuracy of
gauge-invariant quantities because much numerical resolution is wasted
on resolving the gauge reflections.
This is particularly the case when the gauge excitations in question 
are high-frequency modes such as those produced along with the so-called 
`junk radiation' in binary black hole initial data.

There is no discernible difference between the two sets of boundary conditions
as far as constraint violations are concerned, which is what we expect 
because both of them are constraint-preserving.

\begin{figure}
  \plot{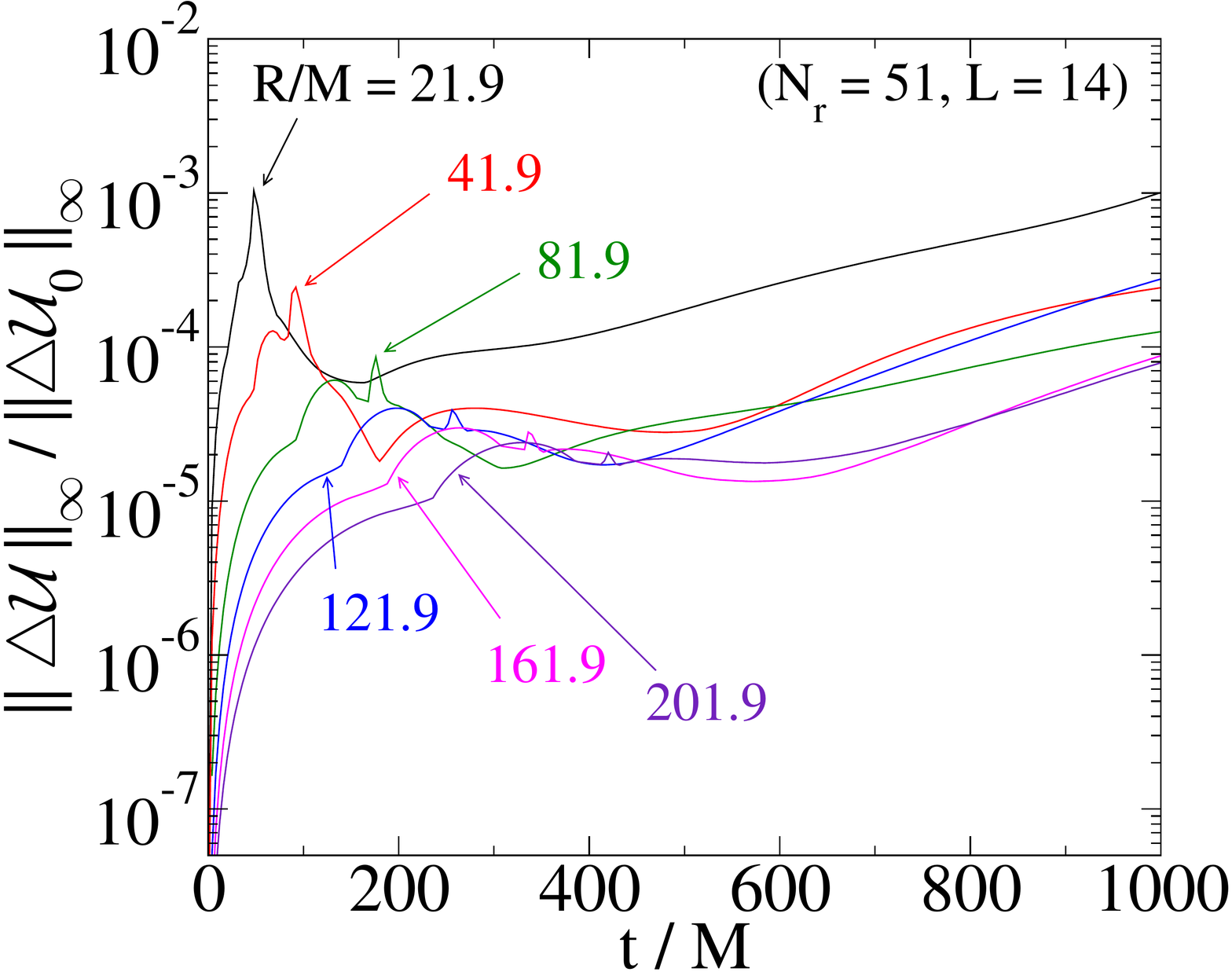}
  \plot{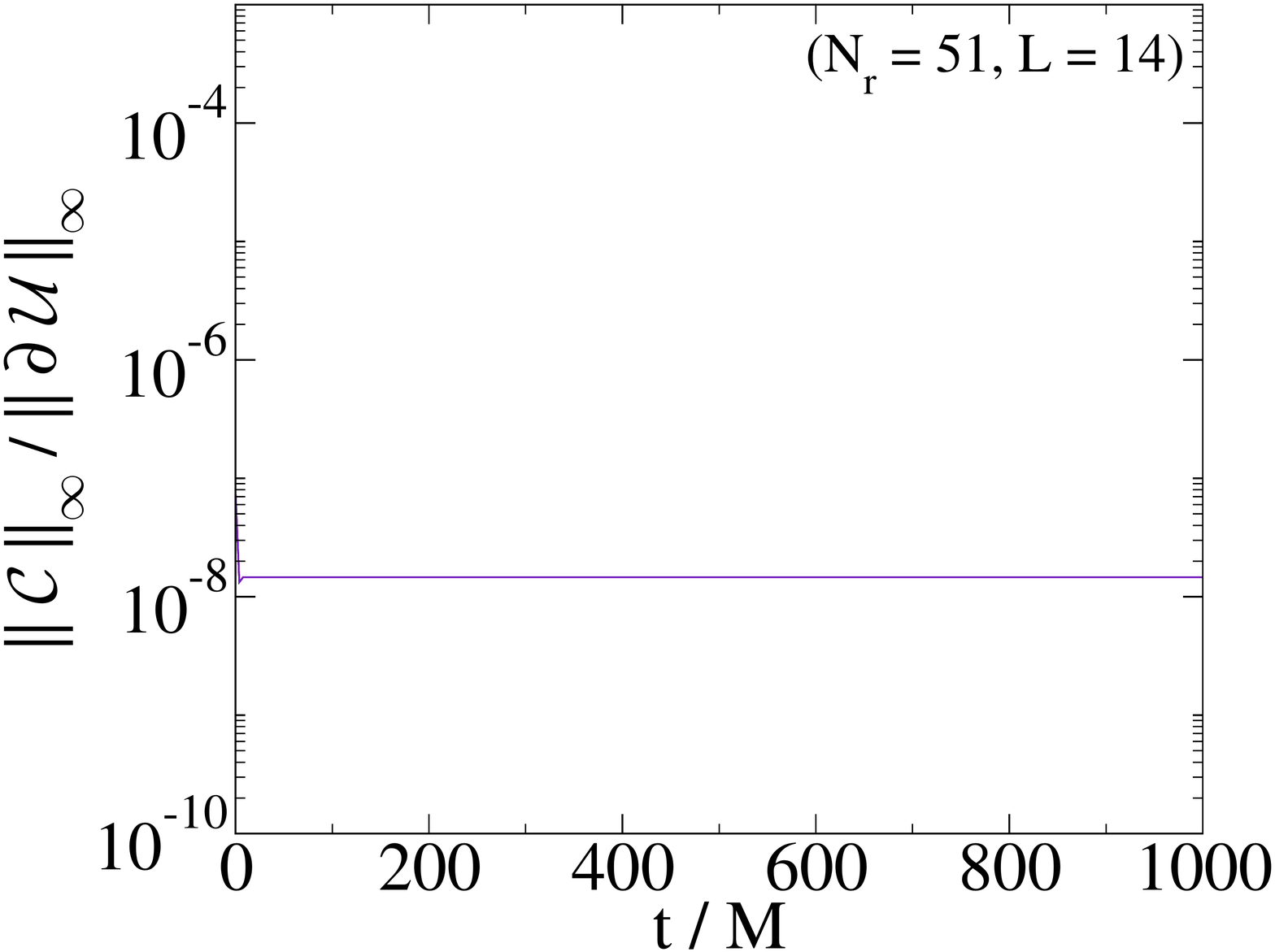}

  \plot{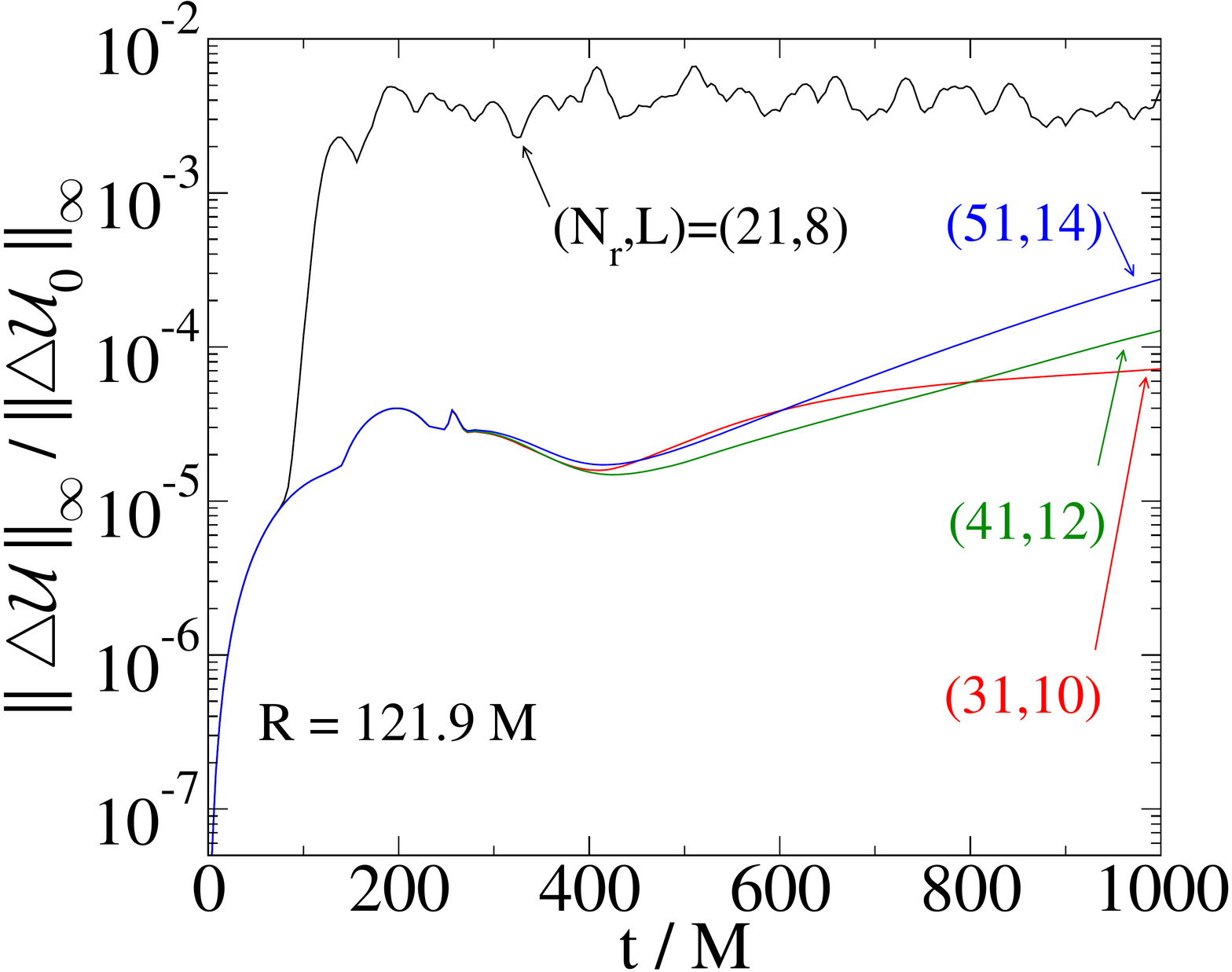}
  \plot{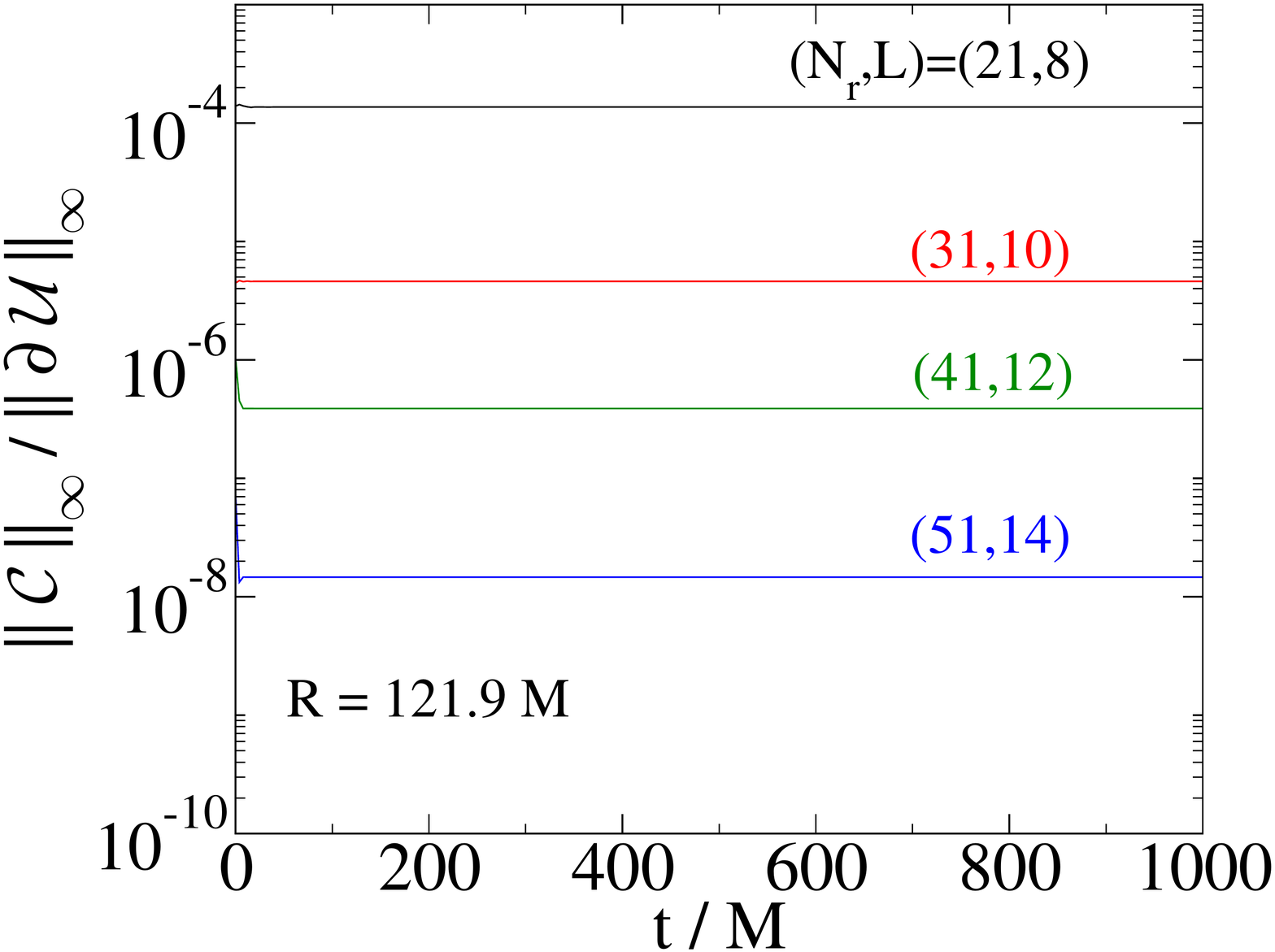}
  \caption{\label{f:NewCpRadii} 
    New CPBCs at different radii. Top half: all radii at the highest
    resolution, bottom half: $R = 121.9 M$ at all resolutions.
    In the top right panel, curves for all outer boundary radii coincide.
  }
\end{figure}

We close this section by investigating the dependence of the
reflections on the radius of the outer boundary (figure \ref{f:NewCpRadii}).
The amplitude of the first peak in $||\Delta\U||_\infty$ decreases as the
boundary is moved outward, roughly like $1/R$.
At late times, there appears to be a power-law growth of that quantity
at a rate that increases slightly with resolution. 
Inspection of the constraints (also in figure \ref{f:NewCpRadii}) 
and $\Psi_4$ (figure \ref{f:DeltaPsi4}) suggests that this is a pure 
gauge effect. 
This blow-up is completely dominated by the innermost 
domain, which contains a long-wavelength feature that is growing in time. 
We speculate that this problem might be cured by a more clever choice of 
gauge source function close to the black hole horizon.

\section{Alternate boundary conditions}
\label{s:AltBC}

In this section, we consider several alternate boundary conditions
that are often used in numerical relativity.
All of these are local conditions imposed at a finite boundary radius,
then in section \ref{s:AltApproach} we consider some additional non-local
boundary treatments.
We run the alternate boundary conditions on our test problem and compare 
the results with the CPBCs 
(using the new gauge boundary condition \eref{e:NewGaugeBC}).

\subsection{Freezing the incoming fields}
\label{s:Freezing}

A very simple boundary condition is obtained by freezing in time 
\emph{all} the incoming fields at the boundary, i.e., 
\begin{equation}
  \label{e:FreezingBC}
  \partial_t u^{1-}_{ab} \doteq 0 \qquad
  (\mathrm{and} \,\, \partial_t u^2_{Aab} \doteq 0 \,\, 
   \mathrm{if} \,\, N^n \, \dot > \, 0) .
\end{equation}
This boundary condition is attractive from a mathematical point of view
because it is of maximally 
dissipative type and hence, together with the symmetric hyperbolic 
evolution equations \eref{e:ghev}, yields a strongly well-posed 
IBVP \cite{Rauch1985,Secchi1996a,Secchi1996b}.
However, in general this boundary condition is not 
compatible with the constraints.

\begin{figure}
  \plot{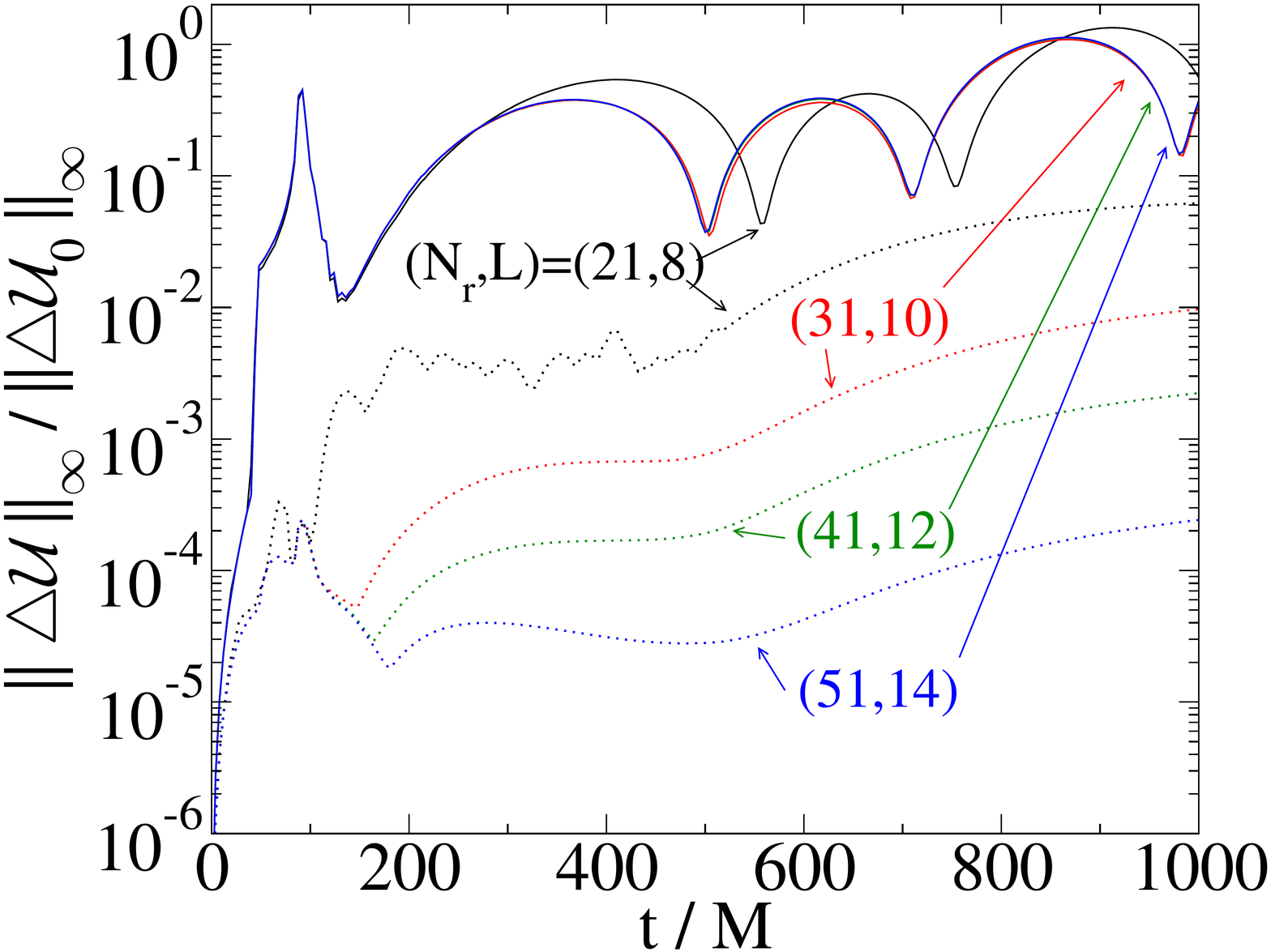}
  \plot{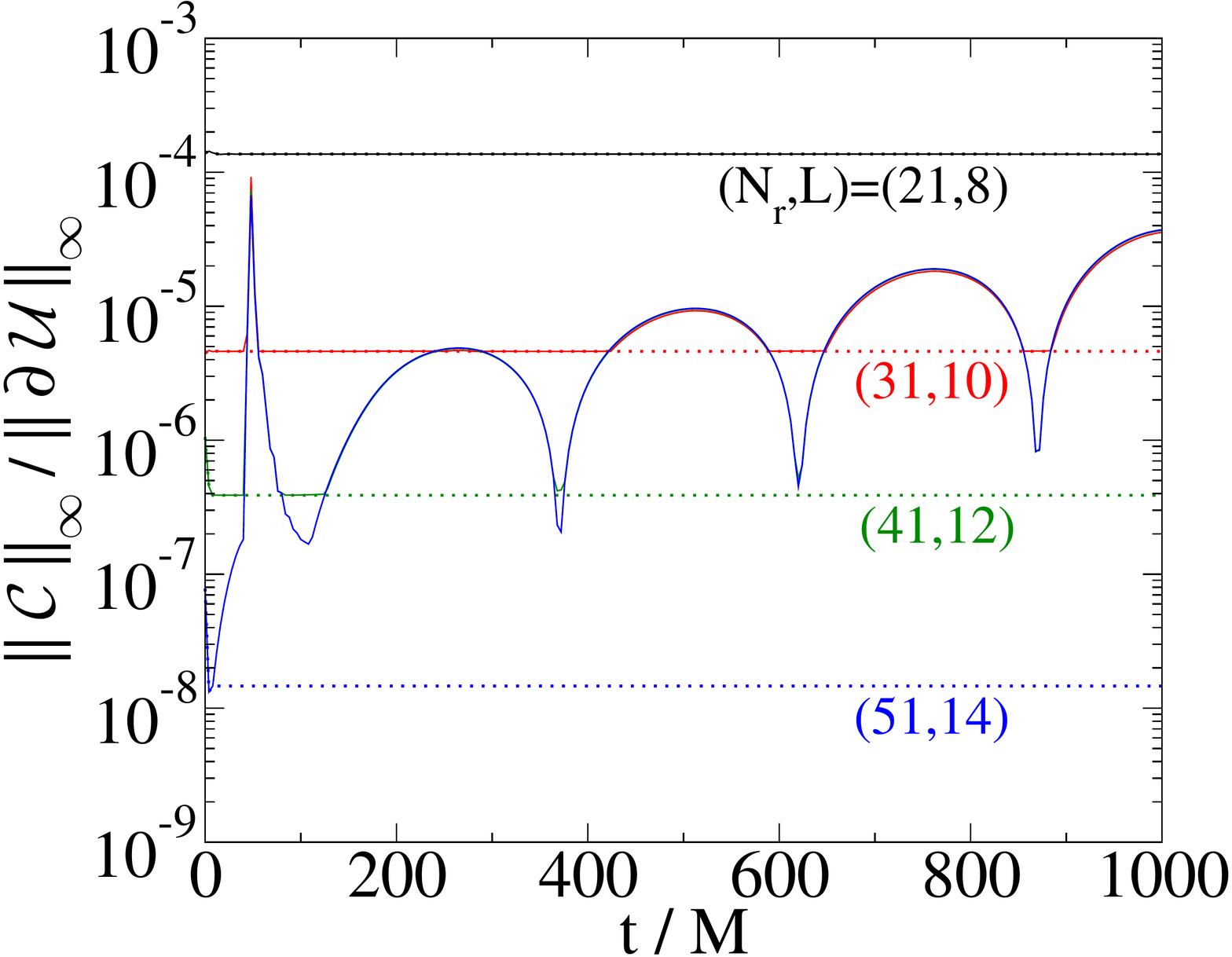}
  \caption{\label{f:FreezingVsNewCp} 
    Freezing (solid) vs. new CPBCs (dotted).
    Four different resolutions are shown: $(N_r, L) = (21, 8)$, 
    $(31, 10)$, $(41, 12)$, and $(51, 14)$. For freezing boundary conditions,
    both $||\Delta\U ||$ and $\C$ converge to a nonzero function with increasing
    resolution. The outer boundary is at $R = 41.9 M$.
  }
\end{figure}

The left side of figure \ref{f:FreezingVsNewCp} demonstrates that 
freezing boundary conditions cause a significantly larger 
(by $\approx 3$ orders of magnitude) initial reflection than
our CPBCs. 
The difference with respect to the reference solution remains large 
in the subsequent evolution and unlike for the CPBCs
does not decrease with increasing resolution.
Furthermore, the violations of the constraints (right side of
figure \ref{f:FreezingVsNewCp}) do not converge away. 
This means that a solution to the Einstein
equations is not obtained in the continuum limit.

\subsection{Sommerfeld boundary conditions}
\label{s:Sommerfeld}

A boundary condition that is often imposed in conjunction with the
BSSN \cite{Shibata1995,Baumgarte1998} formulation of the Einstein equations 
is a Sommerfeld condition on all the components of the spatial metric
$g_{ij}$ and extrinsic curvature $K_{ij}$,
\begin{equation}
  \label{e:BSSNSommerfeld}
  (\partial_t + \partial_r + r^{-1}) 
    \left( \begin{array}{c} g_{ij} - \delta_{ij} \\ K_{ij} \end{array} \right)
    \doteq 0.
\end{equation}
This condition has been used for example in many recent binary black hole 
simulations~\cite{Brugmann2004,Campanelli2006,Baker2006, Diener2006, 
Herrmann2006}.
We cannot impose precisely the conditions \eref{e:BSSNSommerfeld}
in our simulations because there is no one-to-one relationship
between $g_{ij}$ and $K_{ij}$, and the incoming characteristic fields of our
generalized harmonic formulation of Einstein's equations. Instead we
consider the similar condition
\begin{equation}
  \label{e:SommerfeldBC1}
  (\partial_t + \partial_r + r^{-1}) (\psi_{ab} - \eta_{ab}) \doteq 0
\end{equation}
on all the components of the spacetime metric ($\eta_{ab}$ being the 
Minkowski metric).
A very similar boundary condition (without the $r^{-1}$ term) has recently 
been used in the generalized harmonic evolutions of \cite{Szilagyi2006}.

In our formulation, boundary conditions are required not on the spacetime
metric itself but only on certain combinations of its derivatives.
By taking a time derivative of \eref{e:SommerfeldBC1} and rewriting in 
terms of incoming characteristic fields, we obtain
\begin{equation}
  \label{e:SommerfeldBC}
  \partial_t [u^{1-}_{ab} + (\gamma_2 - r^{-1}) \psi_{ab}] \doteq 0.
\end{equation}
This then is our version of the Sommerfeld boundary condition
(cf.~\eref{e:NewGaugeBC}), to be imposed on a spherical 
boundary in the far field (where linearized theory is assumed to be valid).

Because the BSSN formulations using \eref{e:BSSNSommerfeld} are
usually second-order in space, there is no analogue of our three-index
constraint \eref{e:C3} in that system.  To mimic this situation in our
tests of
equation ~(\ref{e:SommerfeldBC}), we also impose a CPBC
on $u^2_{Aab}$ as discussed in section \ref{s:BCConstr},
which together with our constraint damping terms ensures that
violations of the three-index constraint \eref{e:C3} are exponentially damped.

\begin{figure}
  \plot{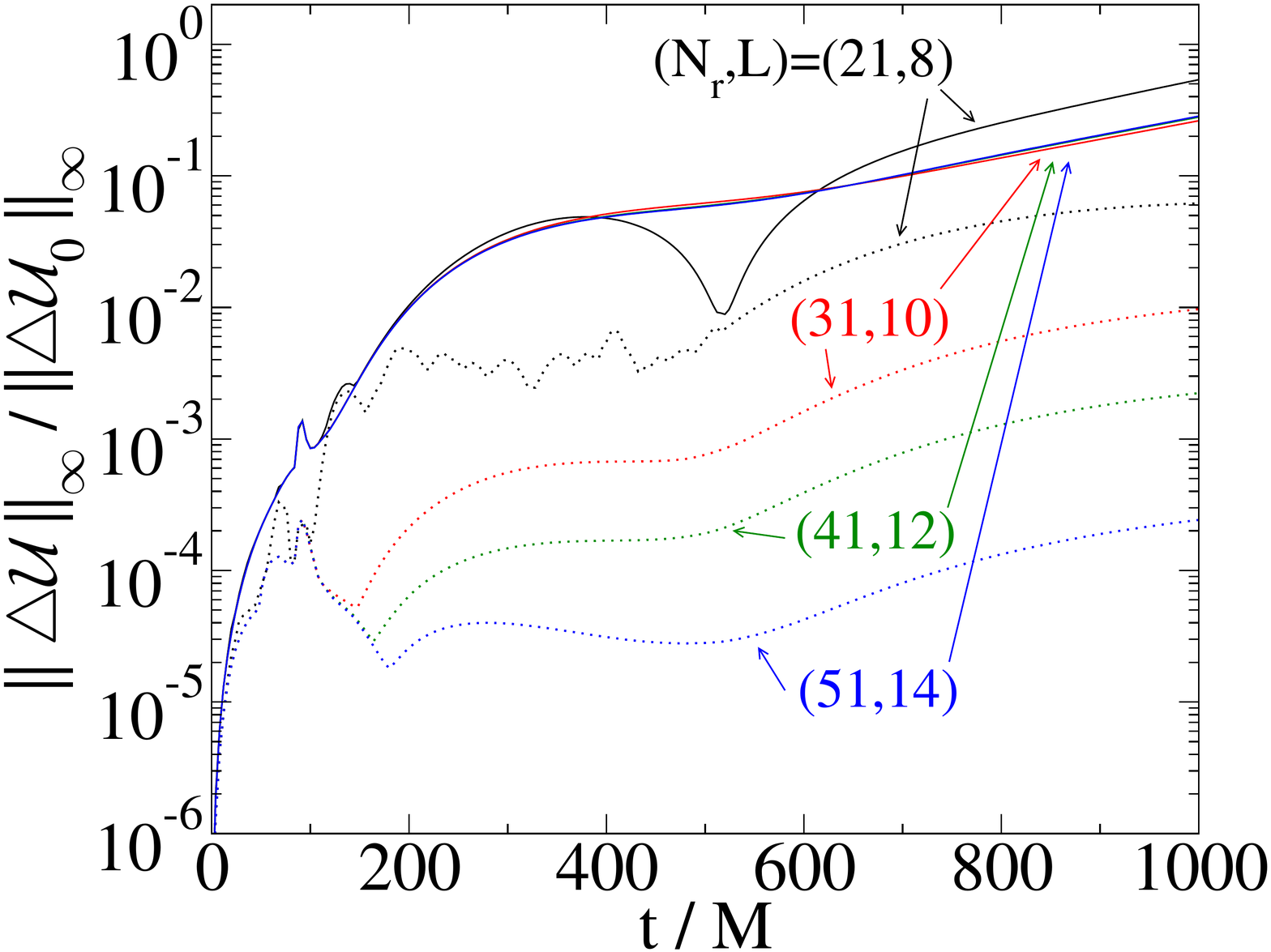}
  \plot{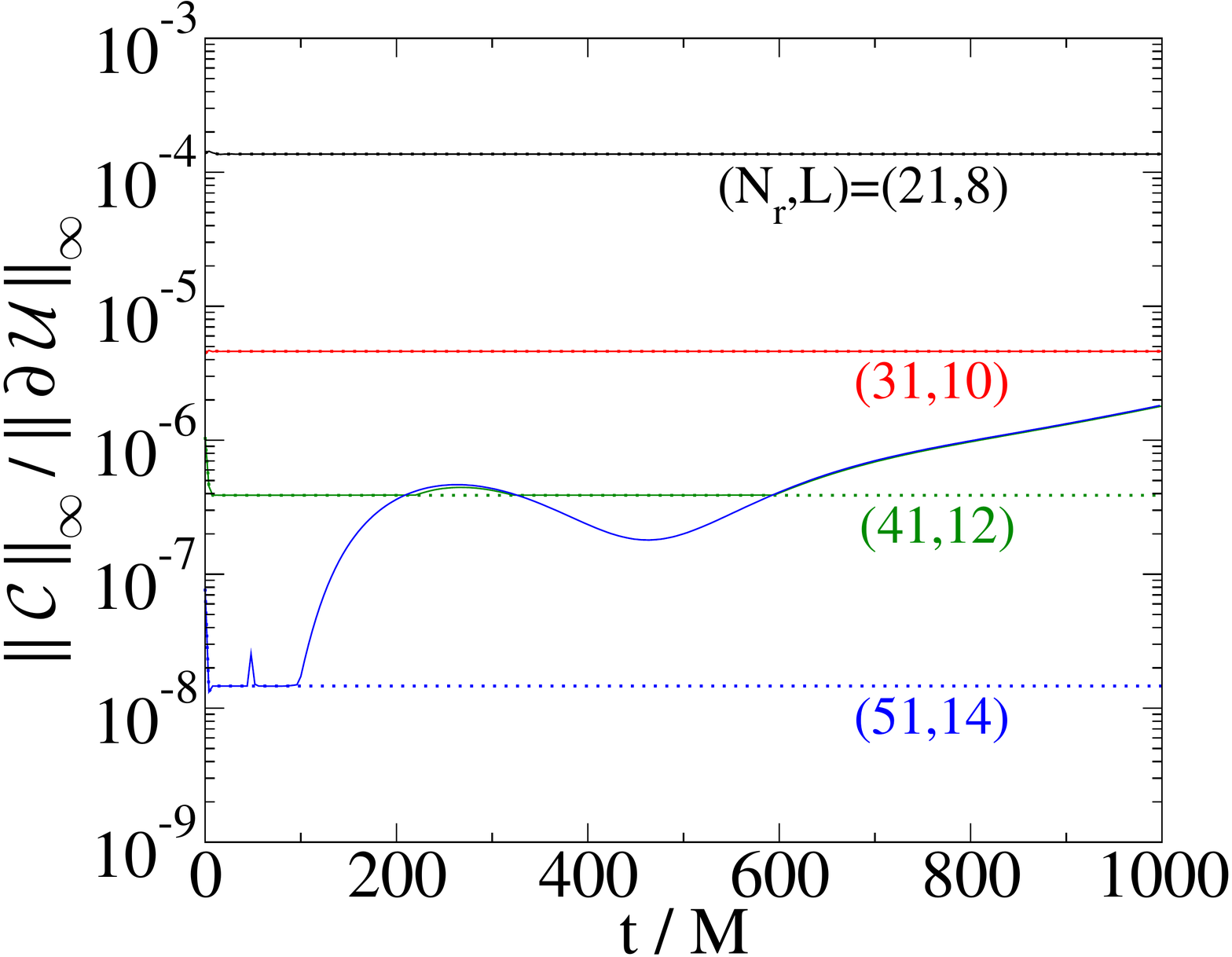}
  \caption{\label{f:SommerfeldVsNewCp} 
    Sommerfeld (solid) vs. new CPBCs (dotted).
    Four different resolutions are shown: $(N_r, L) = (21, 8)$, 
    $(31, 10)$, $(41, 12)$, and $(51, 14)$.
    The outer boundary is at $R = 41.9 M$.
  }  
\end{figure}

Our version of Sommerfeld boundary conditions performs similarly on
our test problem (figure \ref{f:SommerfeldVsNewCp}) 
to the freezing boundary conditions \eref{e:FreezingBC} 
(figure \ref{f:FreezingVsNewCp}).
The initial pulse of reflections is smaller by 
$\approx 2$ orders of magnitude, but later $||\Delta\U ||$ grows to a 
similar level as for freezing boundary conditions. 
Again the constraints do not converge away, although this 
non-convergence appears only at
somewhat higher resolutions than in the freezing case. 

\subsection{Kreiss-Winicour boundary conditions}
\label{s:KreissWinicour}

Recently, Kreiss and Winicour \cite{Kreiss2006} proposed a set of
`Sommerfeld-like' CPBCs for the
harmonic Einstein equations and showed that they result in an IBVP
that is well-posed in the generalized sense. Their boundary conditions
were implemented and tested in \cite{Babiuc2007}; here we compare their
performance with the various other boundary treatments.

The Kreiss-Winicour boundary conditions are obtained by requiring the
harmonic constraint to vanish at the boundary,
\begin{equation}
  \label{e:ZeroC1}
  \C_a \doteq 0.
\end{equation}
In our notation, this can be written as an algebraic condition on 
part of the incoming fields $u^{1-}$,
\begin{equation}
  \label{e:KreissWinicourBC1}
  P^{\mathrm{C'}\,cd}_a u^{1-}_{cd} \doteq F_a ,
\end{equation}
where 
\begin{eqnarray}
  \label{e:KreissWinicourDetails}
  P^{\mathrm{C'}\,cd}_a &=& \textstyle \frac{\sqrt{2}}{4} 
    [2 k^{(c} \delta_a{}^{d)} - k_a \psi^{cd}] , \nonumber\\
  F_a &=& \textstyle \frac{\sqrt{2}}{2} l^b u^{1+}_{ab} 
    - \textstyle \frac{\sqrt{2}}{4} l_a \psi^{bc} u^{1+}_{bc}
    + P^{ij} u^2_{ija} - \half P_a{}^i \psi^{bc} u^2_{ibc} \\
    && - \gamma_2 t_a + H_a . \nonumber
\end{eqnarray}
The range of the projection operator $P^\mathrm{C'}$ is
identical with that of $P^\mathrm{C}$ defined in \eref{e:CPBC}.
For the unconstrained incoming fields $\tilde u^{1-}$ (i.e.~$u^{1-}$
without the $\gamma_2$ term, equation \eref{e:u1tilde}), 
Kreiss and Winicour
\cite{Kreiss2006} specify certain free boundary data $q^\mathrm{P}_{ab}$ 
and $q^\mathrm{G}_{ab}$. In our notation,
\begin{equation}
  \label{e:KreissWinicourBC2}
  P^{\mathrm{P}\,cd}_{ab} \tilde u^{1-}_{cd} = q^\mathrm{P}_{ab}, \qquad
  P^{\mathrm{G}\,cd}_{ab} \tilde u^{1-}_{cd} = q^\mathrm{G}_{ab} .
\end{equation}
In the linearized wave and gauge wave tests of \cite{Babiuc2007},
these boundary data are obtained from the known exact solutions. 
In the absence of an exact solution, it
is suggested that the data could be obtained from an exterior
Cauchy-characteristic or Cauchy-perturbative code. However, since we
do not have such an exterior code, we compute the boundary data from
the background solution, i.e.~Schwarzschild spacetime.
As in the Sommerfeld case (section \ref{s:Sommerfeld}), we use a
constraint-preserving boundary condition on $u^2_{Aab}$ to
emulate the second-order formulation of \cite{Babiuc2007,Kreiss2006},
and this value of $u^2_{Aab}$ is then used to compute $F_a$ in 
\eref{e:KreissWinicourDetails}.

\begin{figure}
  \plot{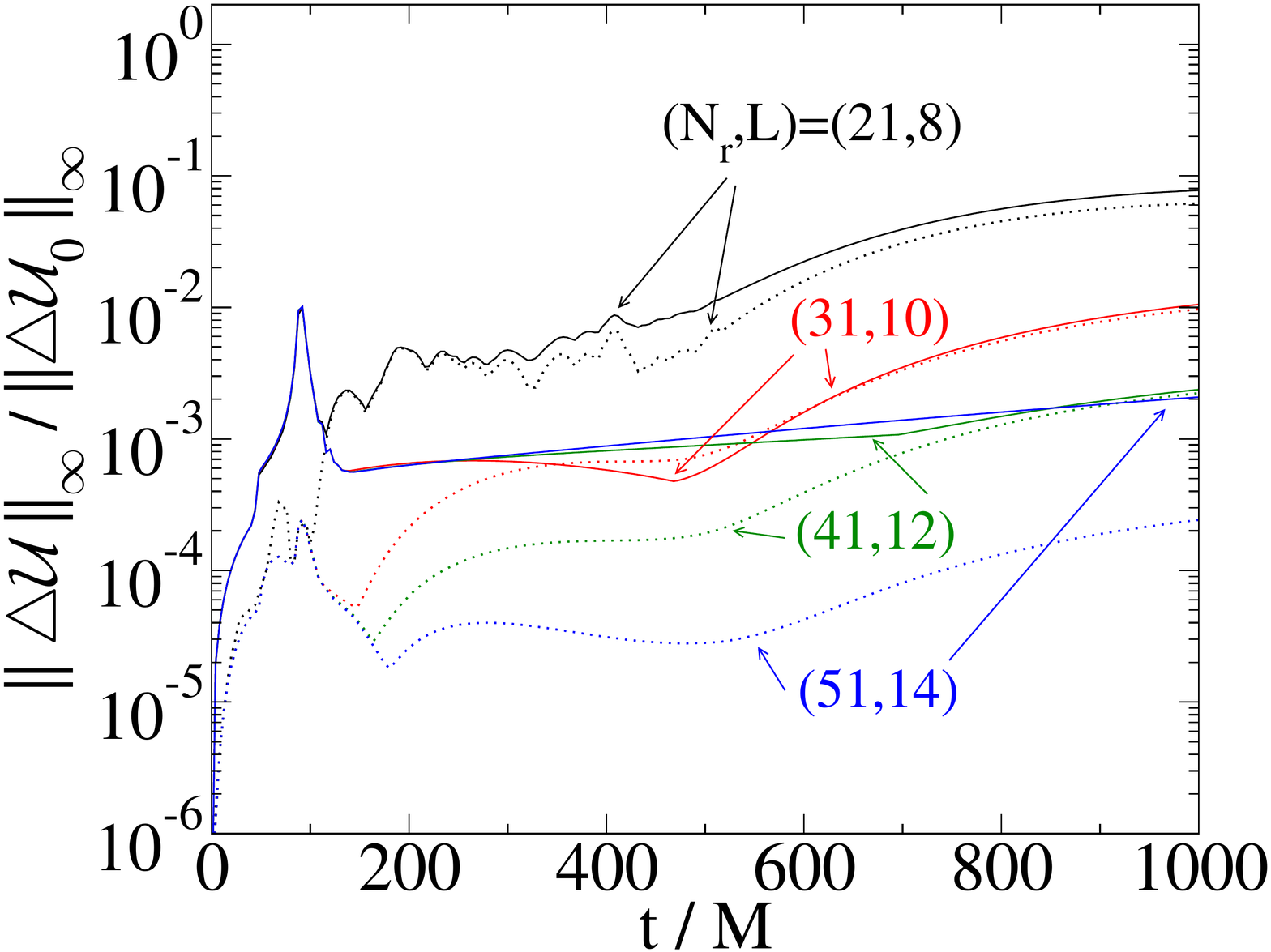}
  \plot{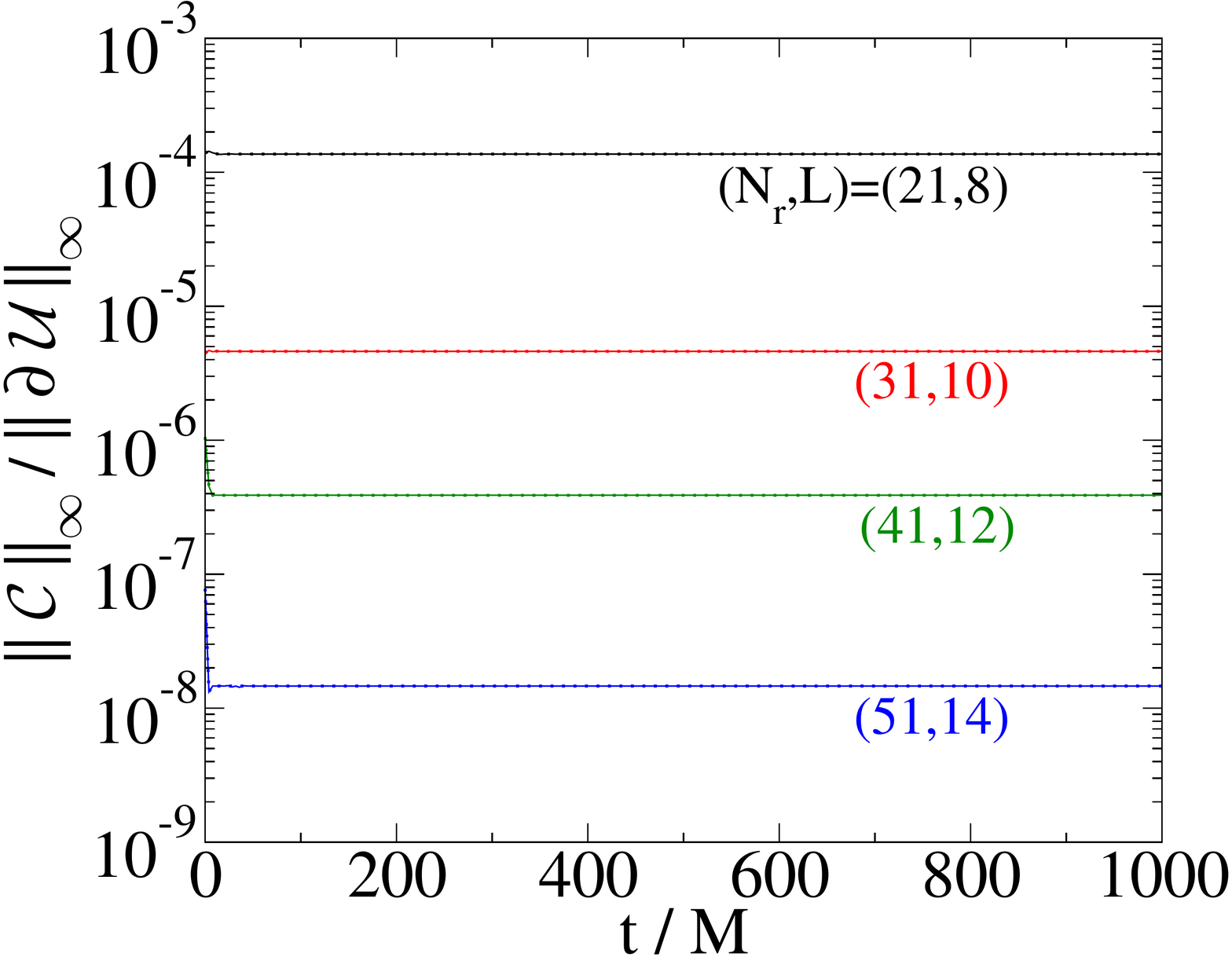}
  \caption{\label{f:KreissWinicourVsNewCp} 
    Kreiss-Winicour (solid) vs. new CPBCs (dotted).
    Four different resolutions are shown: $(N_r, L) = (21, 8)$, 
    $(31, 10)$, $(41, 12)$, and $(51, 14)$.
    The outer boundary is at $R = 41.9 M$.
  }
\end{figure}

Figure \ref{f:KreissWinicourVsNewCp} shows the numerical results for
our test problem. 
The magnitude of the initial reflections lies between that of freezing
and Sommerfeld boundary conditions and is somewhat smaller at later times, 
though still larger than for our CPBCs at the higher resolutions.
The constraints converge away with increasing resolution, as they
should for a boundary condition that is consistent with the constraints.
In a numerical simulation, violations of the constraints are in
general present in the interior of the computational domain.  
These propagate as described by the constraint evolution system 
(\ref{e:ConstraintEvolution}) and some may hit the outer boundary.
The Dirichlet boundary conditions \eref{e:ZeroC1} might be expected to 
cause more reflections of constraint violations than our no-incoming-field 
conditions \eref{e:CPBC}, however, no indications of this are seen in 
figure \ref{f:KreissWinicourVsNewCp}. Probably the constraint damping we use is
sufficiently effective in eliminating the source of these reflections.

We shall see in section \ref{s:AccuracyRadiation} that the
Kreiss-Winicour boundary conditions also cause larger errors
in
the physical degrees of freedom than our CPBCs.
Since the main difference between the two sets of boundary conditions is
our use of a physical boundary condition $\partial_t\Psi_0 \doteq 0$, 
we conclude 
that such a condition is crucial in reducing the reflections from the 
outer boundary.

\section{Alternate approaches}
\label{s:AltApproach}

So far we have only considered boundary conditions that are local
algebraic or differential conditions imposed at the boundary
of some finite computational domain. There are of course many
ways of treating the outer boundary that do not fall into that
category. In this section, we evaluate two such approaches: spatial
compactification and sponge layers.

\subsection{Spatial compactification}
\label{s:Compact}

Spatial compactification is a method that has been widely used in 
numerical relativity, for instance in \cite{Garfinkle2001,Choptuik2003}
or more recently in the generalized
harmonic binary black hole simulations of Pretorius 
\cite{Pretorius2005a,Pretorius2005b,Pretorius2006}.

\begin{figure}
  \plot{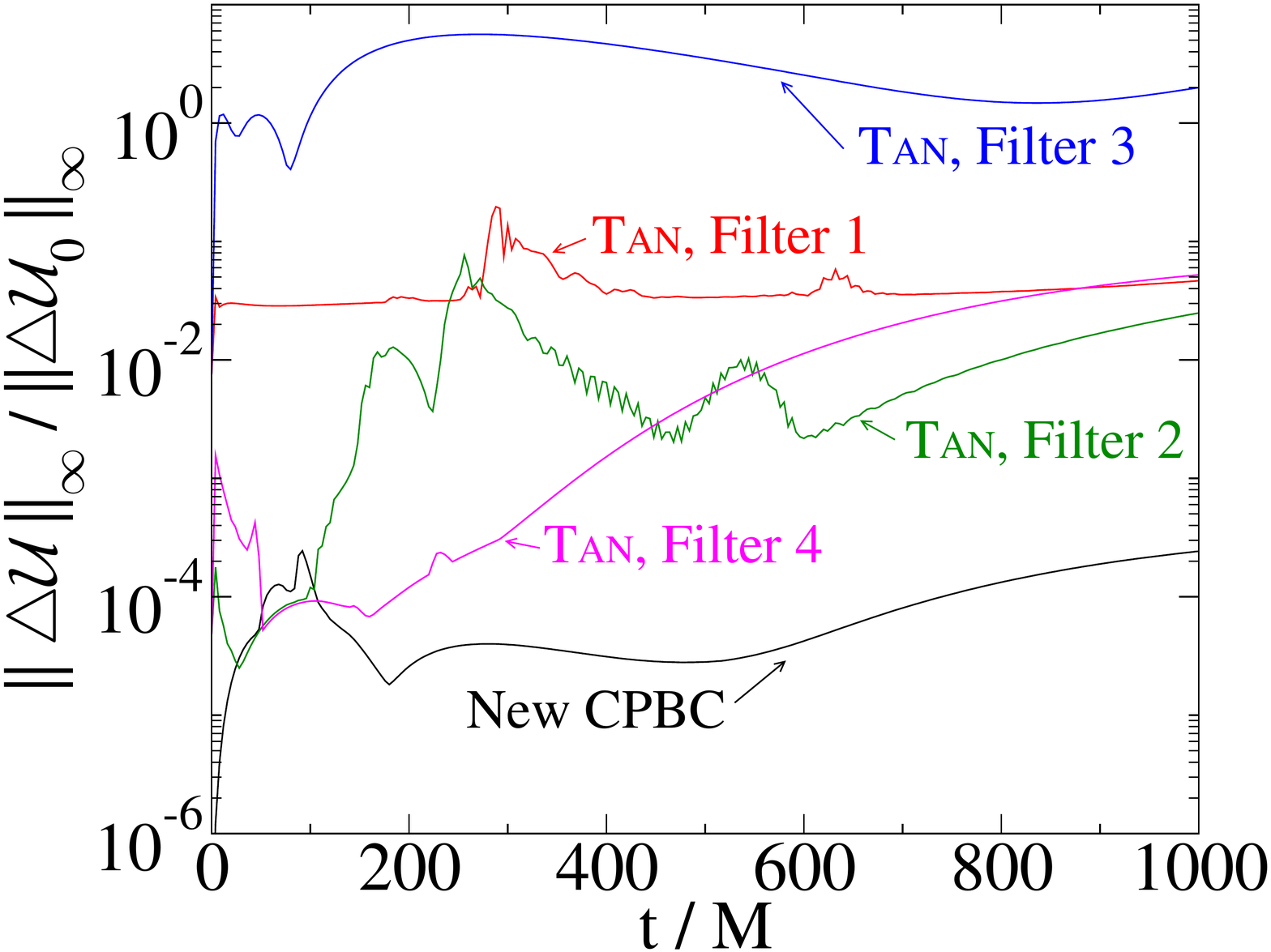}
  \plot{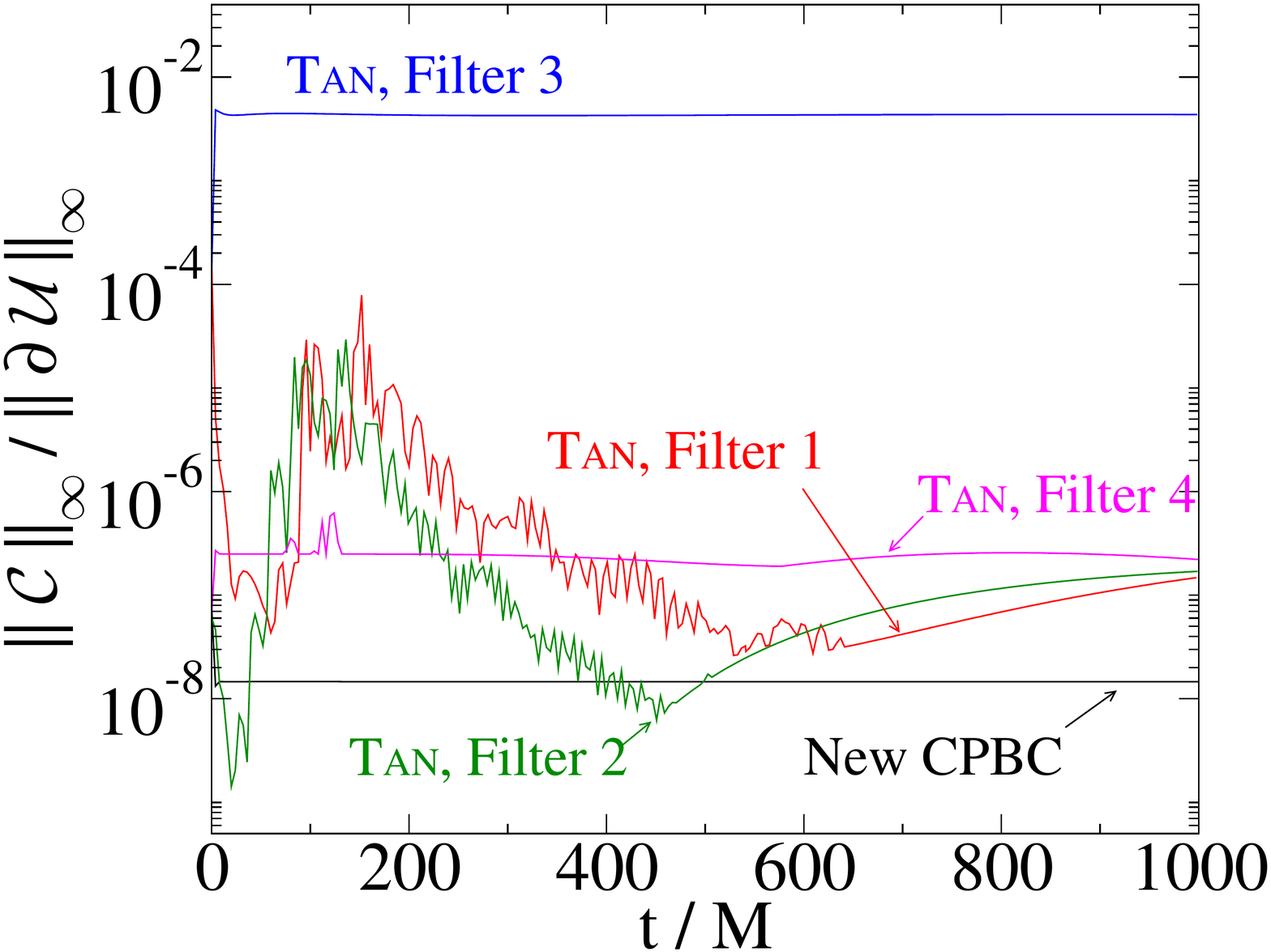}
  \caption{\label{f:TanVsNewCp} 
    \textsc{\small Tan} compactification with various filters vs. new CPBCs.
    Only the highest resolution $(N_r, L) = (51, 14)$ is shown.
    The compactification scale (and the radius of the outer boundary
    in the CPBC case) is $R = 41.9 M$.
  }
\end{figure}

\begin{figure}
  \plot{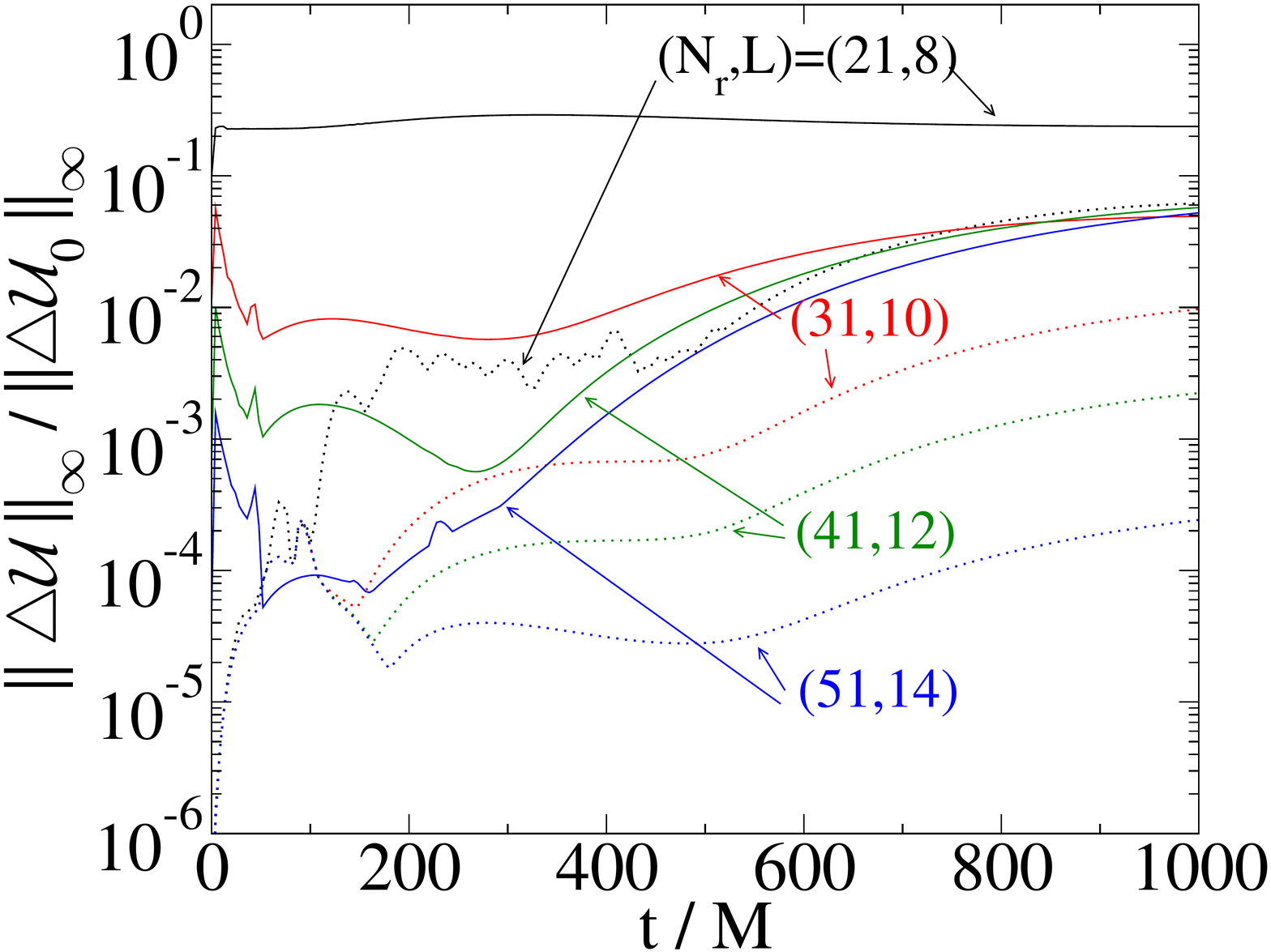}
  \plot{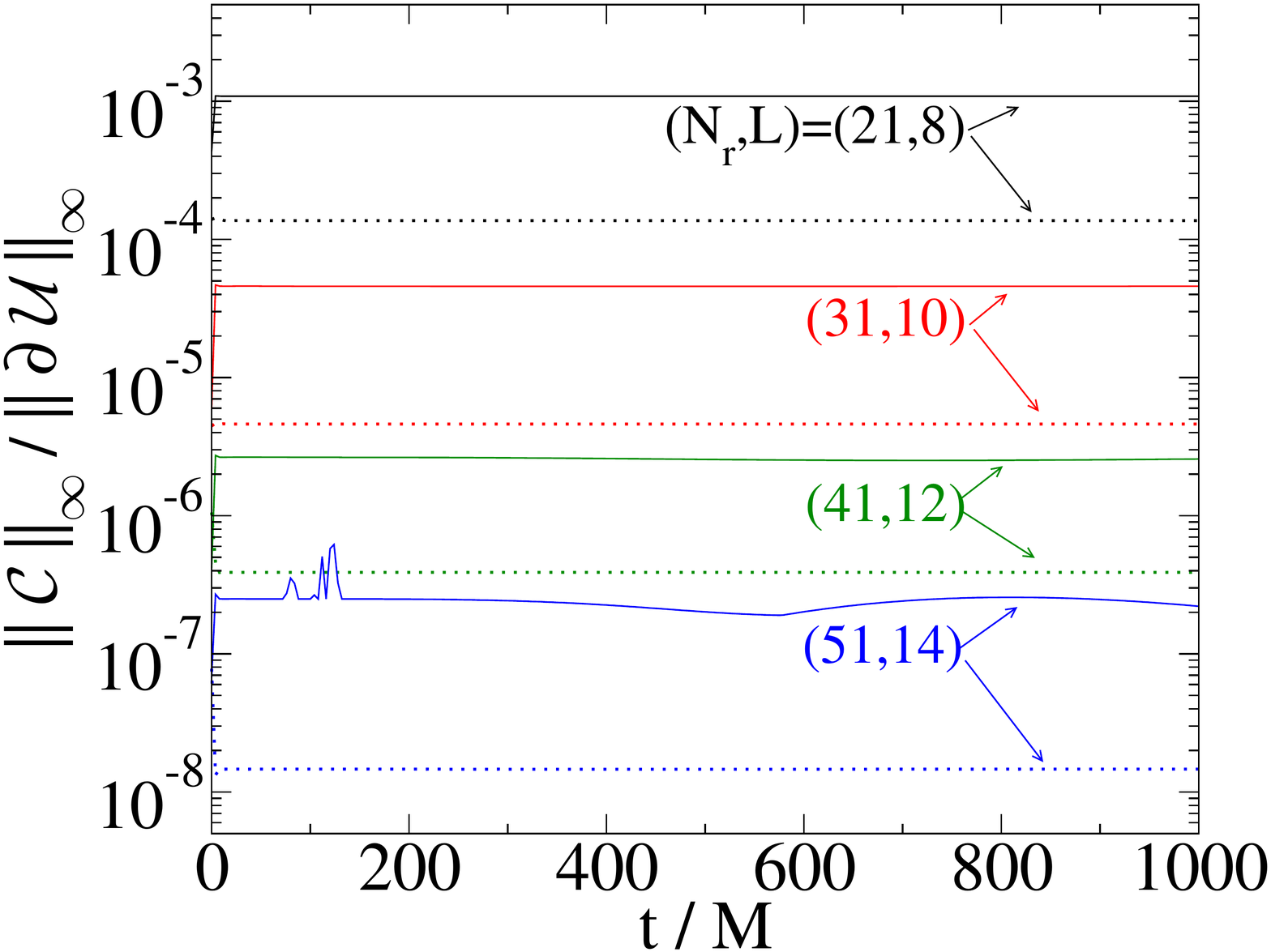}
  \caption{\label{f:BestTanVsNewCp} 
    \textsc{\small Tan} compactification with filter 4 (solid) vs. 
    new CPBCs (dotted).
    Four different resolutions are shown: $(N_r, L) = (21, 8)$, 
    $(31, 10)$, $(41, 12)$, and $(51, 14)$.    
    The compactification scale (and the radius of the outer boundary
    in the CPBC case) is $R = 41.9 M$.
}
\end{figure}

The basic idea is to introduce spatial coordinates
that map spacelike infinity to a finite location. Here we 
consider mappings that are functions of coordinate radius only (whereas
Pretorius applies the mapping to each Cartesian coordinate separately).
We have used two such mappings, named \textsc{\small Tan} and \textsc{\small Inverse}, 
as detailed in  \ref{s:CompactDetails}.  Each map has a scale $R$
across which the mapping is (essentially) linear.
The outermost grid point is placed at a very large but finite uncompactified
radius ($r = 10^{17} M$).  With respect to the compactified radial coordinate, 
the characteristic speeds are below numerical roundoff there and hence no
boundary condition should be needed.  The following results were produced
using constraint-preserving boundary conditions;
we have checked for one simulation that using no boundary condition at
all yields results that are visually indistinguishable from the ones
presented here on the scales of figures \ref{f:TanVsNewCp},
\ref{f:BestTanVsNewCp}, and \ref{f:DeltaPsi4}.

As the waves travel outward, they become more and more blue-shifted with
respect to the computational grid and are eventually no longer
properly resolved. However, some form of artificial numerical 
dissipation is applied that acts as a low-pass filter and causes the
waves to be damped as they become increasingly distorted.
We have experimented with various such filters; see  
\ref{s:CompactDetails} for details. One of them (referred to as number 2 in the
following) is designed to emulate as closely as possible
the fourth-order Kreiss-Oliger dissipation
used by Pretorius.

In the following numerical comparisons, we evaluate the differences
with respect to the reference solution only in the part of the domain
where the compatification map is essentially linear, i.e.~for $r \leqslant R$.
First we compare the various filtering methods at a fixed resolution, 
using the \textsc{\small Tan} compactification mapping (figure
\ref{f:TanVsNewCp}).
The filters that are applied to the right side of the evolution
equations (numbers 1 and 3, cf.~table \ref{t:Filtering}) do somewhat
better than those applied to the solution itself (numbers 2 and 4),
and the \textsc{\small Exponential} filters (numbers 3 and 4) are slightly better than
the \textsc{\small Kreiss-Oliger} filters (numbers 1 and 2). All of them are 
outperformed by the CPBCs (imposed at $r = R$). 
For our closest approximation to the dissipation used by Pretorius (number 2), 
$||\Delta\U ||$ is comparable to constraint-preserving boundary
conditions at the peak of reflections (at $t \approx 2 R$) but becomes larger
by about $2$ orders of magnitude at later times.
The compactification methods also generate considerable constraint
violations.

Next we focus on the best filter (number 4) of the previous test but vary
the resolution (figure \ref{f:BestTanVsNewCp}). 
We do see convergence of $||\Delta\U ||$ initially but the convergence
degrades at later times. This is surprising at first because 
with increasing resolution, the waves travel a longer
distance before they fail to be resolved.  Note however that
the high-frequency filter is applied at each time step, as is done in
the simulations of Pretorius.  For higher resolutions, the time steps 
are smaller because of the CFL condition and the filter is applied 
more often, thus leading to a stronger damping of the waves. 
This may well lead to the observed loss of convergence with increasing 
resolution.
The constraints appear to converge away in this test, 
although from figure \ref{f:BestTanVsNewCp} it appears
that this will not persist for even higher resolutions.

We have also evaluated the \textsc{\small Inverse} mapping described 
in \ref{s:CompactDetails}.  The results are similar, but somewhat worse than 
the \textsc{\small Tan} mapping results shown here.

\subsection{Sponge layers}
\label{s:Sponge}

A method that has been used for a long time in computational
science, in particular for spectral methods 
(see e.g.~section 17.2.3 of \cite{Boyd2001} and references therein), 
involves so-called sponge layers. 
A sponge layer is introduced by modifying the evolution equations according to
\begin{equation}
  \label{e:spongemodification}
  \partial_t u = \ldots - \gamma(r) (u - u_0),
\end{equation}
where $u_0$ refers to the unperturbed background solution (Schwarzschild 
spacetime in our case) and the smooth sponge function $\gamma(r) > 0$ is
large only close to the outer boundary of the computational domain.
(Here we use uncompactified coordinates as in sections \ref{s:CPBC} 
and \ref{s:AltBC}.) In this way, the waves are damped exponentially as 
they approach the outer boundary. Details on our particular choice of
$\gamma(r)$ can be found in \ref{s:SpongeDetails}.

\begin{figure}
  \plot{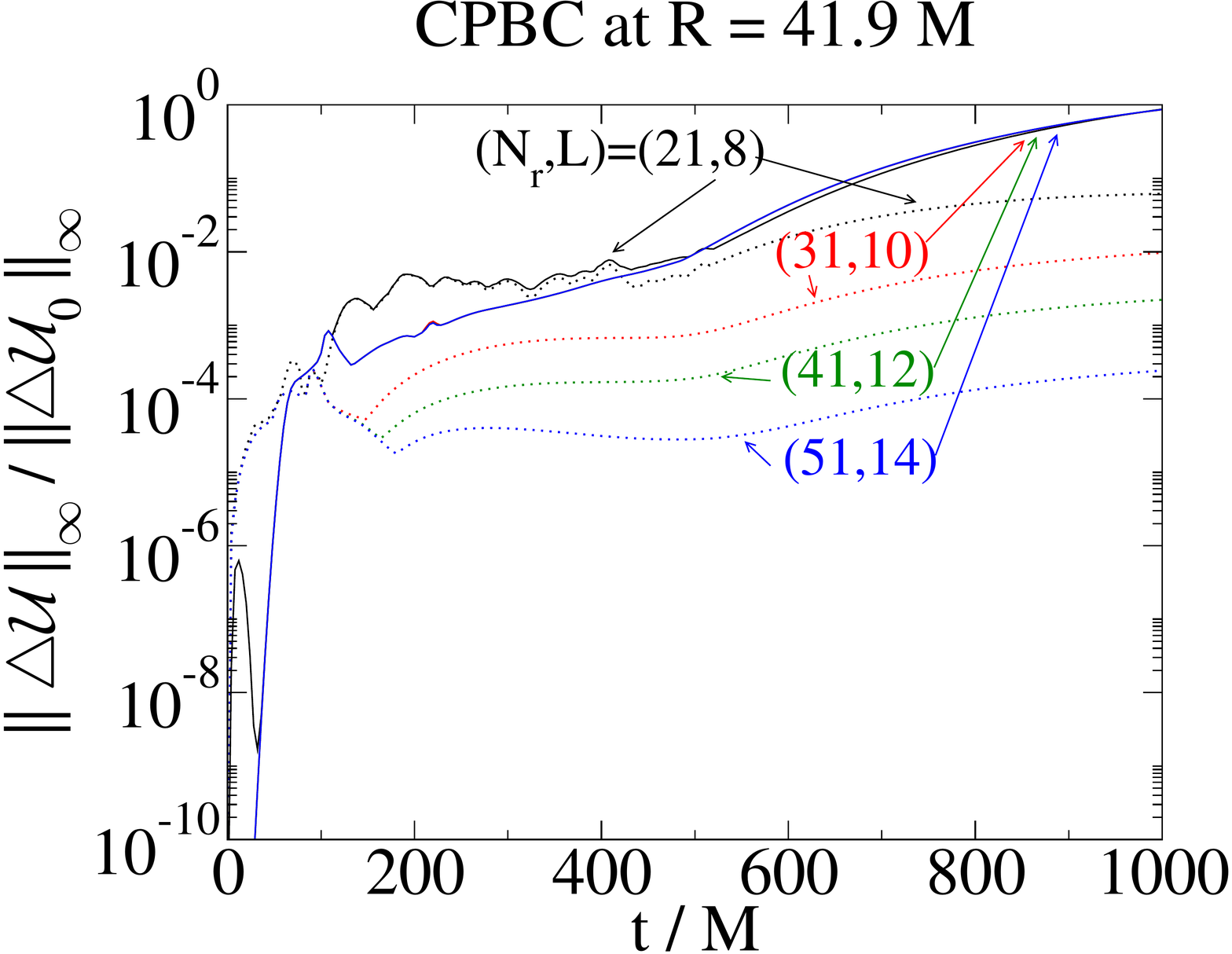}
  \plot{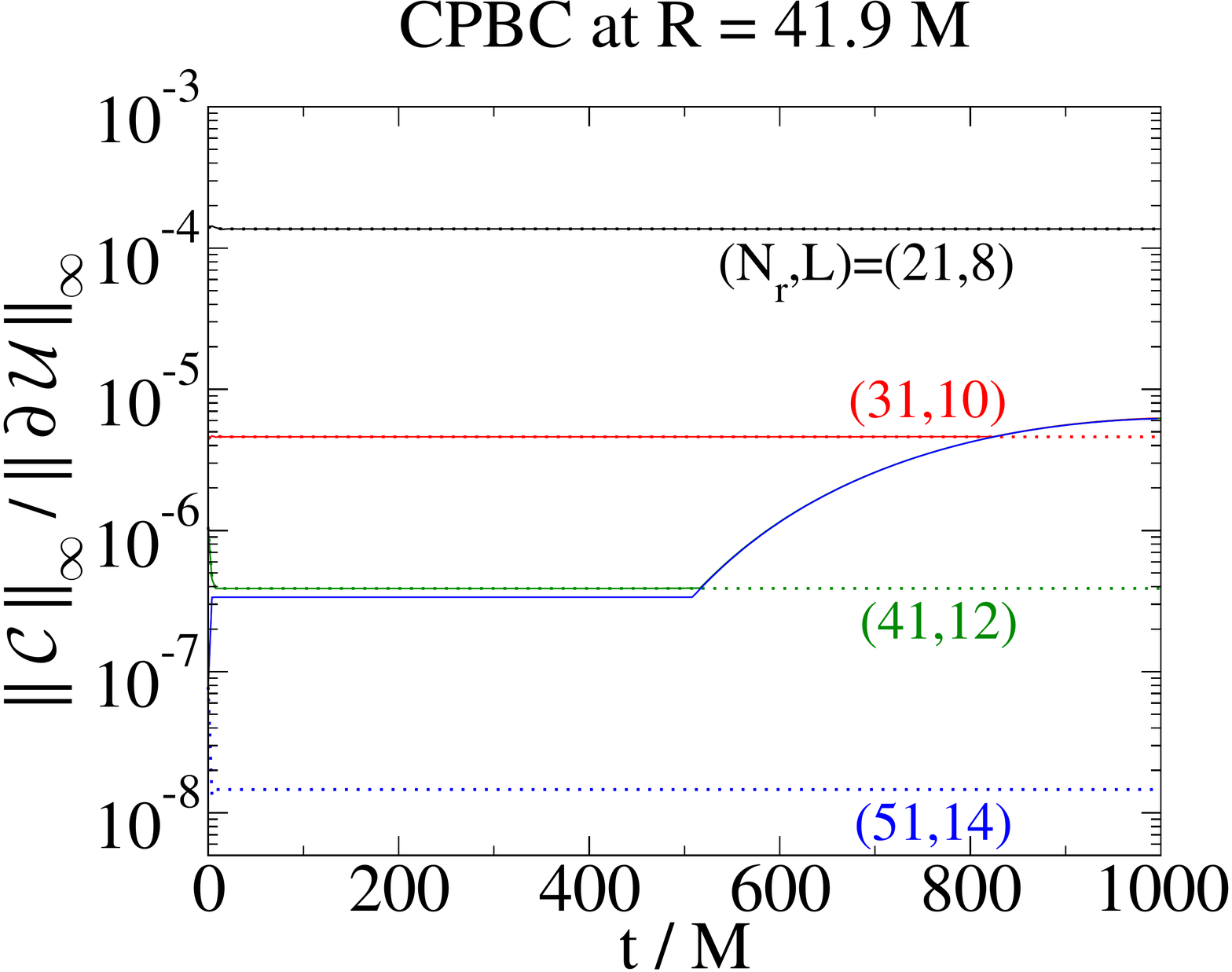}

  \plot{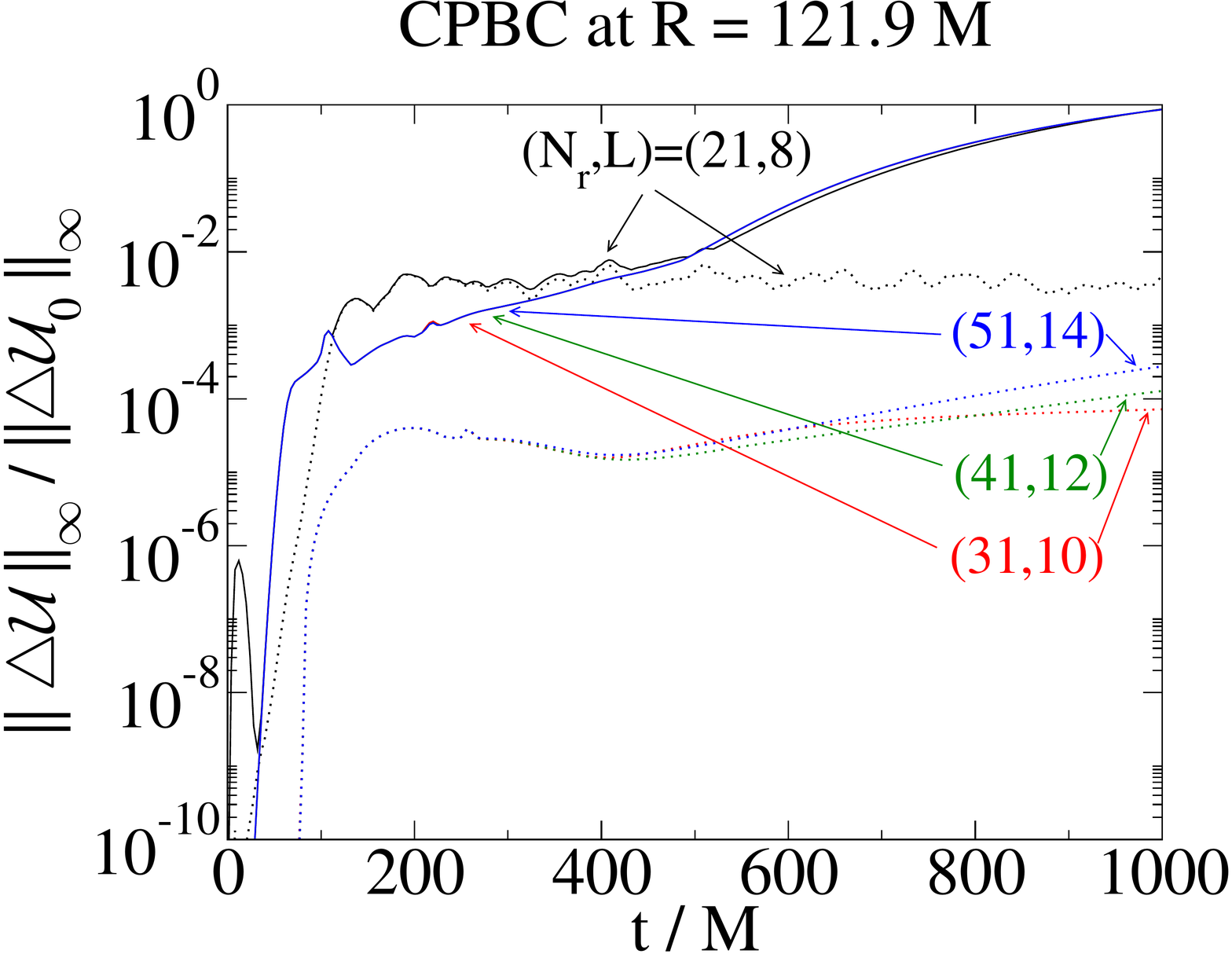}
  \plot{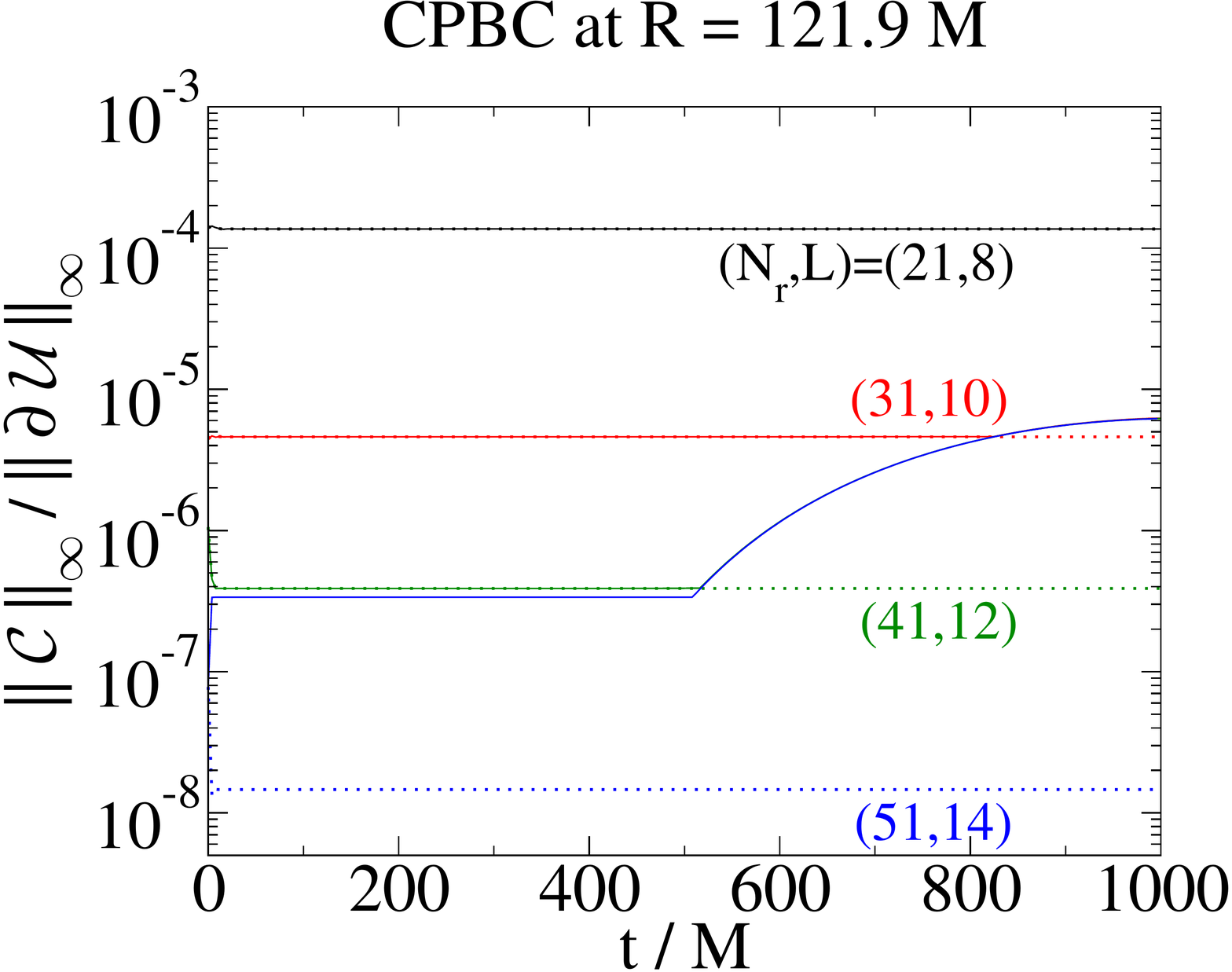}
  \caption{\label{f:SpongeVsNewCp} 
    Sponge layer method (solid) vs. new CPBCs at two different radii (dotted).
    Four different resolutions are shown: $(N_r, L) = (21, 8)$, 
    $(31, 10)$, $(41, 12)$, and $(51, 14)$.    
    The size of the sponge-free region is $R = 41.9 M$ and 
    $||\Delta \U||_\infty$ is only computed for $r \leqslant R$.
}
\end{figure}

We compare the sponge layer method with our CPBCs in 
figure \ref{f:SpongeVsNewCp}. For the CPBCs, the
boundary is either placed at $R = 41.9 M$ (the outer edge of the sponge-free
region) or at $R = 121.9 M$ (the outer edge of the sponge).
At early times ($t \lesssim 2 R$), the $||\Delta \U||_\infty$ of the 
sponge layer method lies between that of the CPBCs
for the two choices of outer boundary radius, 
whereas at later times, it is much larger than both versions of CPBCs.
The constraint violations in the sponge runs do not converge away.

\section{Physical gravitational waves}
\label{s:Radiation}

Perhaps the most important predictions of numerical relativity
simulations at the present time are the gravitational waveforms
produced by astrophysical systems like binary black holes.
It is important therefore to understand how the accuracy of these waveforms 
is affected by the choice of boundary treatment.
Physical gravitational radiation can be described by the
Newman-Penrose scalars $\Psi_4$ and $\Psi_0$.
The scalar $\Psi_4$ is dominated by the outgoing radiation (its
ingoing part is suppressed by a factor of $(kr)^4$, where $k$ is the
wavenumber), whereas $\Psi_0$ is dominated by the ingoing radiation
(its outgoing part is suppressed by a factor of $(kr)^4$).
In this section we compare the gravitational waves extracted 
from the various boundary treatment solutions 
on a sphere of radius $r = R_\mathrm{ex}$, using the methods described in 
\ref{s:WaveExtraction}.

We note that $\Psi_4$ ($\Psi_0$) has a coordinate-invariant
meaning only in the limit as future (past) null infinity is approached.  
The quantities computed at finite radius $r$ will differ from
those observed at infinity by terms of the order $\Or(1/r)$.
In the particular case of perturbed Schwarzschild spacetime considered
here, a gauge-invariant wave extraction method does exist even at
finite radius (see e.g.~\cite{Pazos2006} and references therein) 
but we do not adopt it here.
Our purpose in this paper is merely to measure the effects on $\Psi_4$ 
caused by the various boundary treatments.

\subsection{Difference of $\Psi_4$ with respect to the reference solution}
\label{s:AccuracyRadiation}

\begin{figure}
  \plot{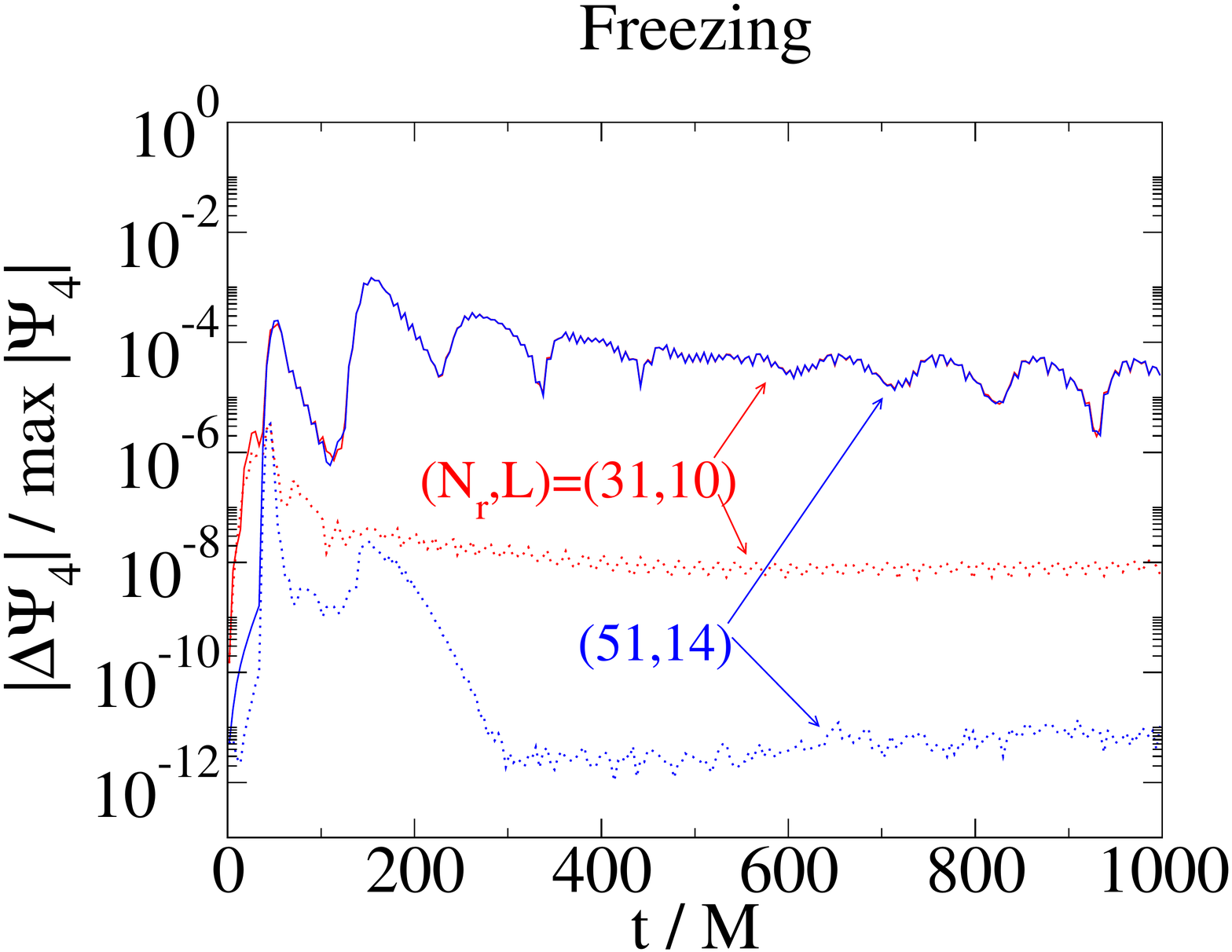}
  \plot{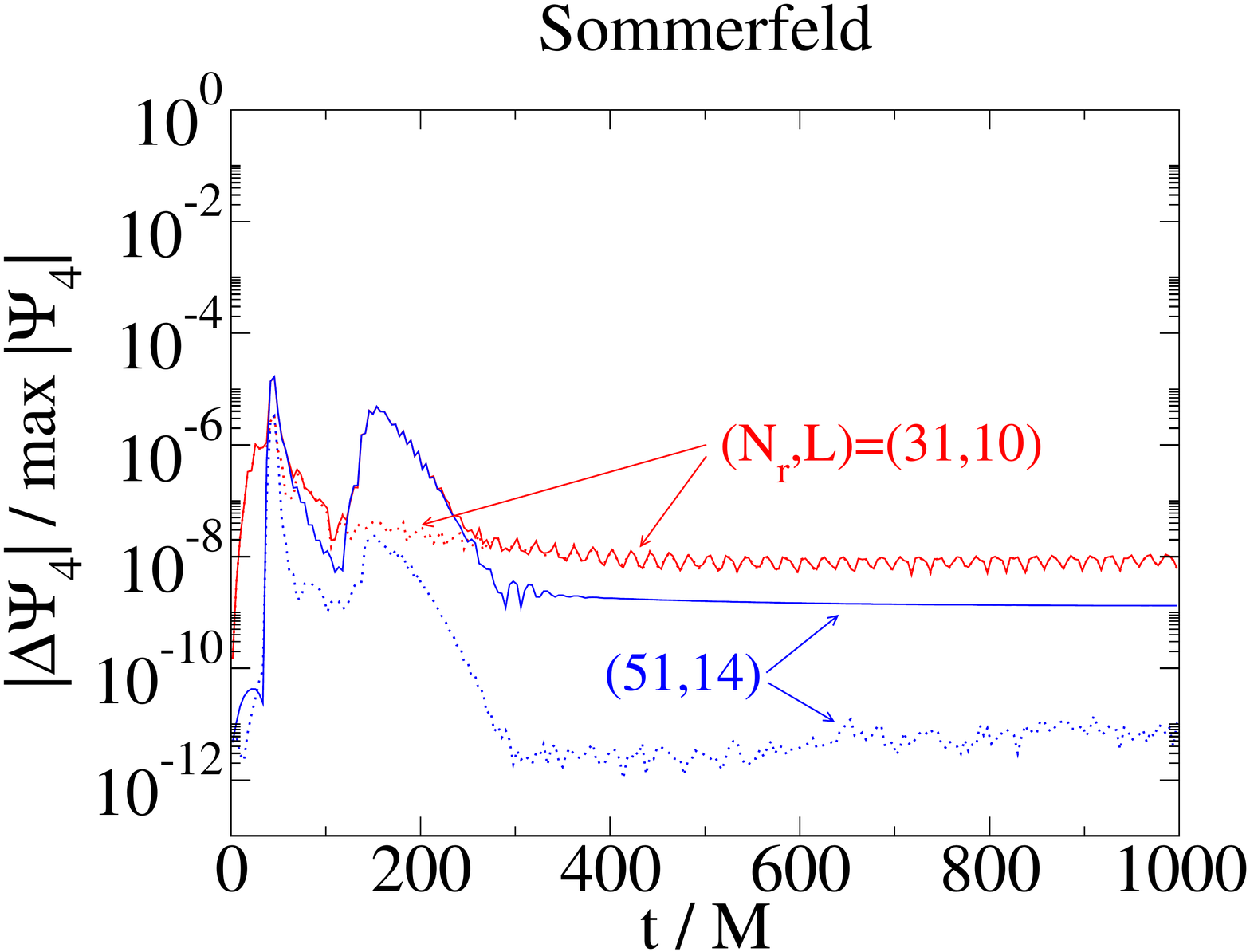}

  \plot{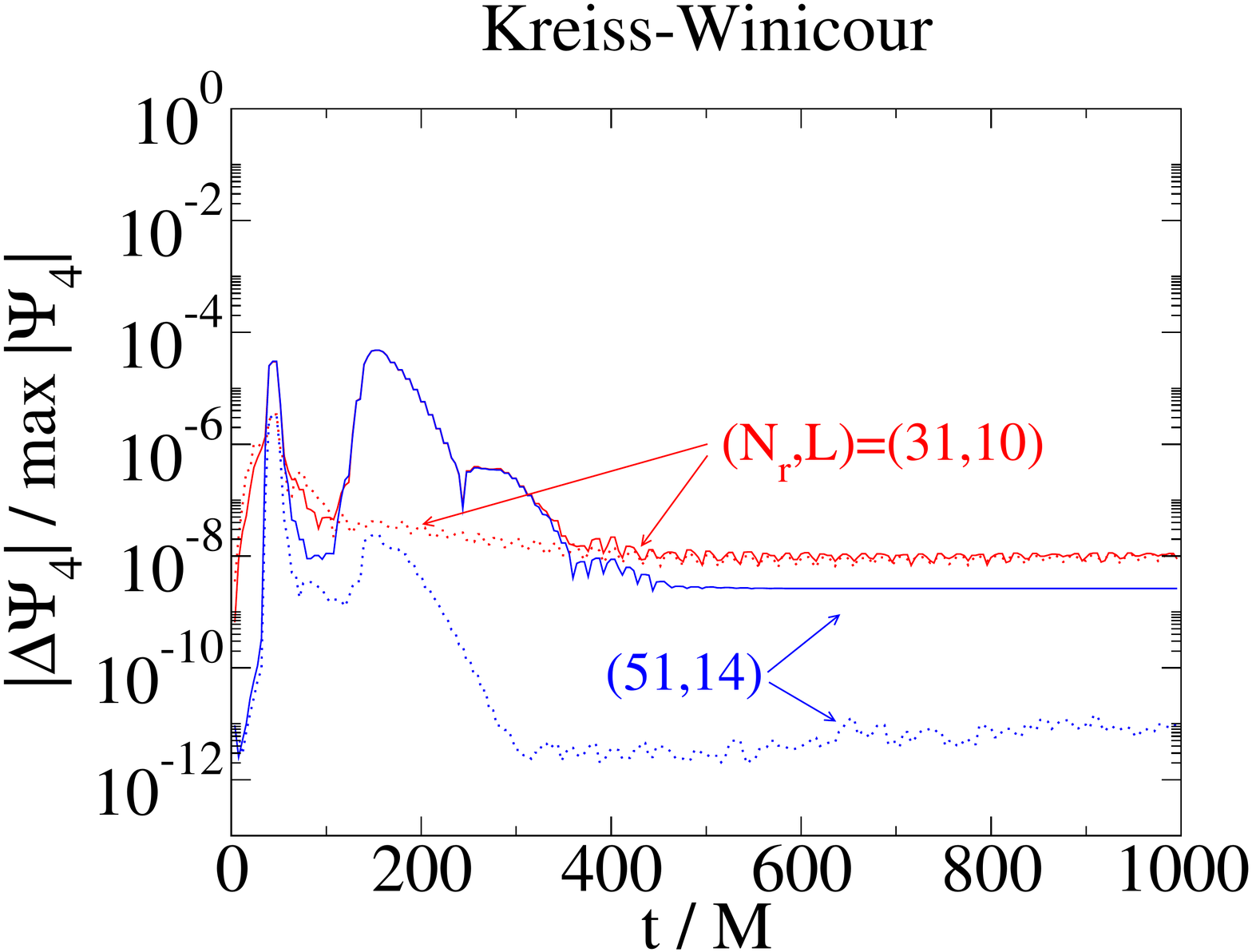}
  \plot{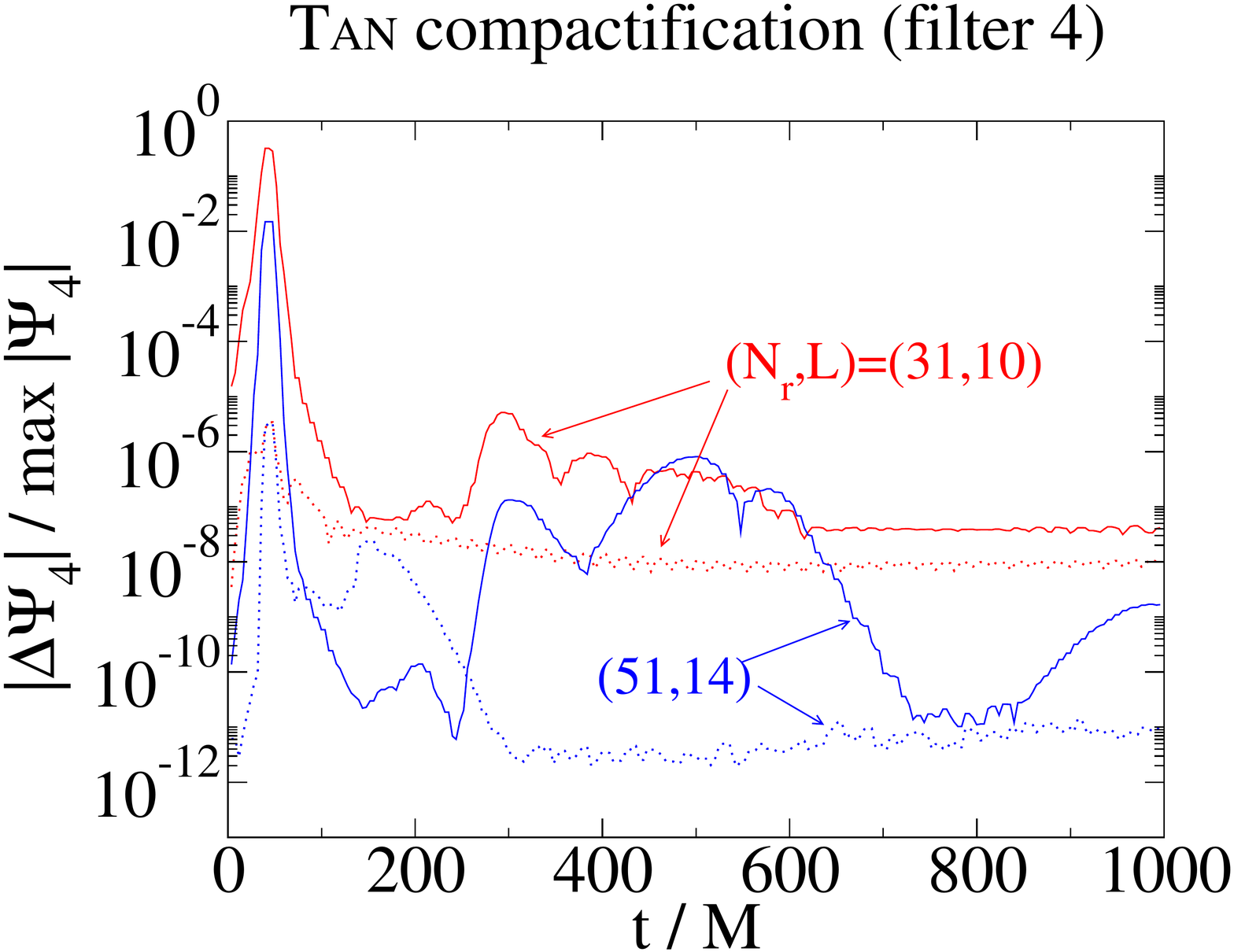}

  \centering \plot{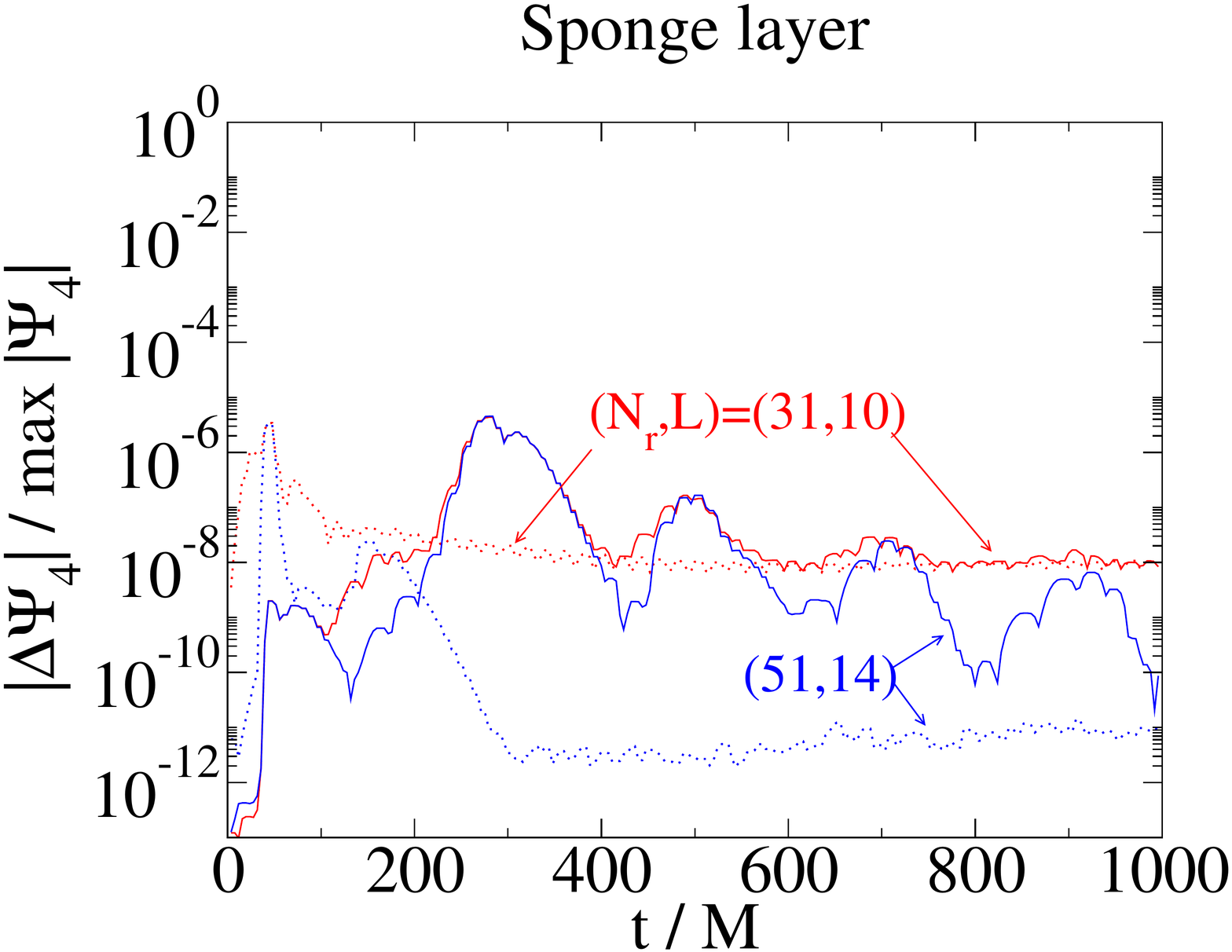}
  \caption{\label{f:DeltaPsi4} 
    Difference of $\Psi_4$ for the various alternate methods (solid) 
    vs. the new CPBCs (dotted).
    Two resolutions are shown: $(N_r, L) = (31, 10)$ and 
    $(51, 14)$.
    The radius of the outer boundary (or the compactification scale, 
    or the size of the sponge-free region, respectively) is $R = 41.9 M$
    and the waves are extracted at $R_\mathrm{ex} = 40 M$.
}
\end{figure}

We begin by evaluating $\Delta\Psi_4 \equiv \Psi_4 - \Psi_4^{\mathrm{ref}}$, 
where $\Psi_4$ is the Newman-Penrose scalar computed using one of the various
boundary methods and $\Psi_4^{\mathrm{ref}}$ is the same quantity computed from
the reference solution at the same extraction radius.
The curves shown in figure \ref{f:DeltaPsi4} plot the maximum value of 
$|\Delta\Psi_4|$ over time intervals of length 20$M$ (this time filtering 
averages over the high frequency quasi-normal oscillations of the 
black hole), normalized by the maximum value of $|\Delta\Psi_4|$ over
the entire evolution.
The radius of the outer boundary (or the compactification scale, 
or the size of the sponge-free region, respectively) used for 
these comparisons is $R = 41.9 M$, and the radiation is extracted nearby
at $R_\mathrm{ex} = 40 M$.  

The first peak in $|\Delta \Psi_4|$ seen in figure \ref{f:DeltaPsi4} 
arises as the wave in our test problem passes outward
through the extraction sphere at $t \approx R_\mathrm{ex}$. This peak
is caused by a presently unknown (probably gauge) interaction between the outer
boundary (or compactified region etc.) and the 
spacetime near the extraction sphere.
We have verified that this interaction and its influence on the peak 
in $\Delta\Psi_4$ goes away if we move the outer boundary (or the extraction
surface) so that they are not in causal contact as the outgoing wave pulse
passes the extraction surface. 

Some of the outgoing radiation is reflected off the boundary.
Most of this reflected radiation is subsequently absorbed by the black hole, 
but some of it excites the hole, which then emits quasi-normal mode
radiation of exponentially decaying amplitude.
This exponential decay can be clearly seen for most of the boundary
treatments.

In the case of freezing boundary conditions, nearly all of the
outgoing quasi-normal mode radiation is reflected from the boundary
because the reflection coefficient is nearly $1$ for the wave number 
of the dominant mode, $k = 0.376 / M$ (cf.~figure \ref{f:ReflCoeff}).
It then re-excites the black hole, which again radiates and so forth.
On average the amplitude of the reflections remains roughly
constant in time for this case.
This behaviour is consistent with the result shown in figure 3 of 
\cite{Lindblom2006} for a similar perturbed black hole simulation. 

For the Sommerfeld and Kreiss-Winicour boundary conditions, the
reflections are much smaller but still considerably larger (by
$2$ to $3$ orders of magnitude) than for our CPBCs.
We attribute this difference largely to our use of the physical boundary 
condition \eref{e:ZeroPsi0}.

The spatial compactification method has the largest difference 
$|\Delta\Psi_4|$,
particularly at early times $t \sim R$ (about $4$ orders of magnitude 
larger than for the CPBCs). 
We suspect that this may be a consequence of the use of artificial dissipation,
as discussed in section \ref{s:Compact}.

The sponge layer method has the smallest errors at early times.
This is not surprising because the outer boundary of the sponge layer is
much further out at $R = 121.9 M$.   However at later times when the
waves begin to interact with the sponge layer, 
this method causes reflections comparable in amplitude to those using
Sommerfeld boundary conditions.

We also note that at late times the level of $|\Delta \Psi_4|$
decreases significantly with resolution for the 
CPBCs, but not generally for the other boundary treatments.

We think it is remarkable that the maximum relative error in the
extracted physical radiation is quite small ($10^{-5}$ to $10^{-3}$)
in these tests, even for the less sophisticated boundary treatments
such as the freezing or Sommerfeld boundary conditions.  This success
is due in part to the fact that the extraction radius, 
$R_{\mathrm{ex}}=40M$, for this test problem is about ten wavelengths 
(of the initial radiation pulse) away from the central black hole.  
Our results are likely to be more accurate than those from typical 
binary black hole simulations, which place the outer boundary at two
or three wavelengths.
This suggests that current binary black hole codes using, for
instance, Sommerfeld boundary conditions, can still produce waveforms 
that are useful for some aspects of gravitational wave data analysis
provided the outer boundary is placed sufficiently far out.
Data analysis applications needing high precision waveforms, however, 
such as source parameter measurement or high-amplitude supermassive binary
black hole signal subtraction for LISA, will need to use a more
sophisticated boundary treatment that produces smaller errors in $\Psi_4$.

\subsection{Comparison with the predicted reflection coefficient}
\label{s:PredictedReflection}

Buchman and Sarbach~\cite{Buchman2006,Buchman2007} have recently developed a
hierarchy of increasingly absorbing physical boundary conditions for 
the Einstein equations by analyzing the equations describing the evolution
of the Weyl curvature on both a flat and a Schwarzschild background spacetime.
Their analysis predicts, in particular, the reflection coefficient
$\rho$ (defined as the ratio of the ingoing to the outgoing parts of the
solution) that arises from the $\partial_t\Psi_0 \doteq 0$ physical
boundary condition that we use.  

For quadrupolar radiation (as in our numerical tests), this reflection 
coefficient is given by equation (89) of \cite{Buchman2006},
\begin{equation}
  \rho(k R) = \textstyle \frac{3}{2} (k R)^{-4} + O(k R)^{-5},
\end{equation}
where $k$ is the wave number of the gravitational radiation and $R$ is
the boundary radius.  
(As explained at the beginning of section \ref{s:GaugeBC}, 
we assume the background spacetime to be flat; effects due to the
backscattering would only enter at $\Or(M/R)$.)
By evaluating $\Psi_0$ and $\Psi_4$ at the
extraction radius of our test, we find that the ratio $\Psi_0 /
\Psi_4$ agrees with their predicted $\rho$ to leading order in
$1/(kR)$.  We note that the tetrad we use for wave extraction
(\ref{s:WaveExtraction}) does not agree exactly with that of
\cite{Buchman2006}.  
However, the tetrads do agree for the unperturbed Schwarzschild solution, 
so that the errors introduced into $\Psi_0$ and $\Psi_4$ due to our
different choice of tetrad are second-order small in perturbation
theory and hence the comparison with \cite{Buchman2006} is consistent.

\begin{figure}
  \plot{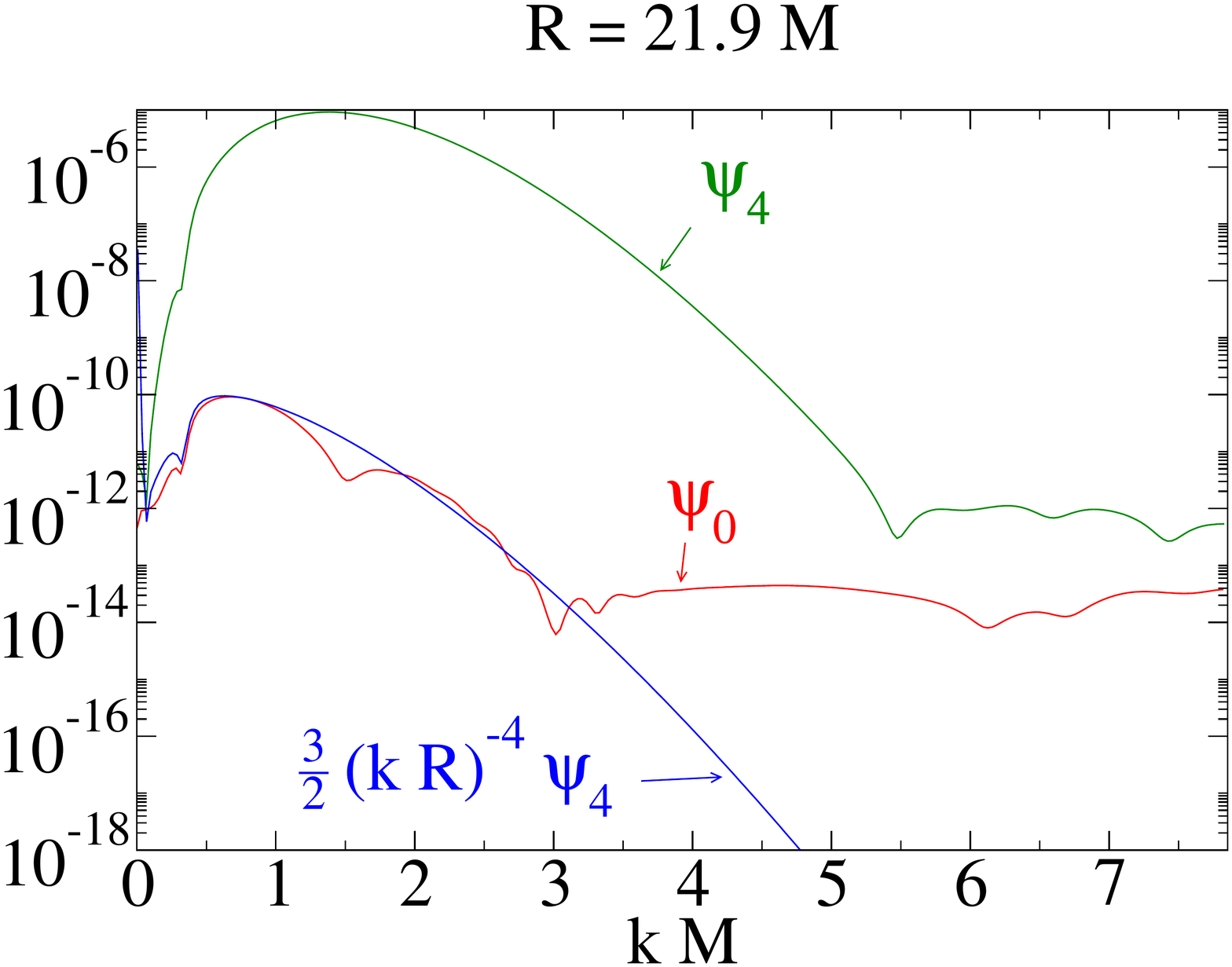}
  \plot{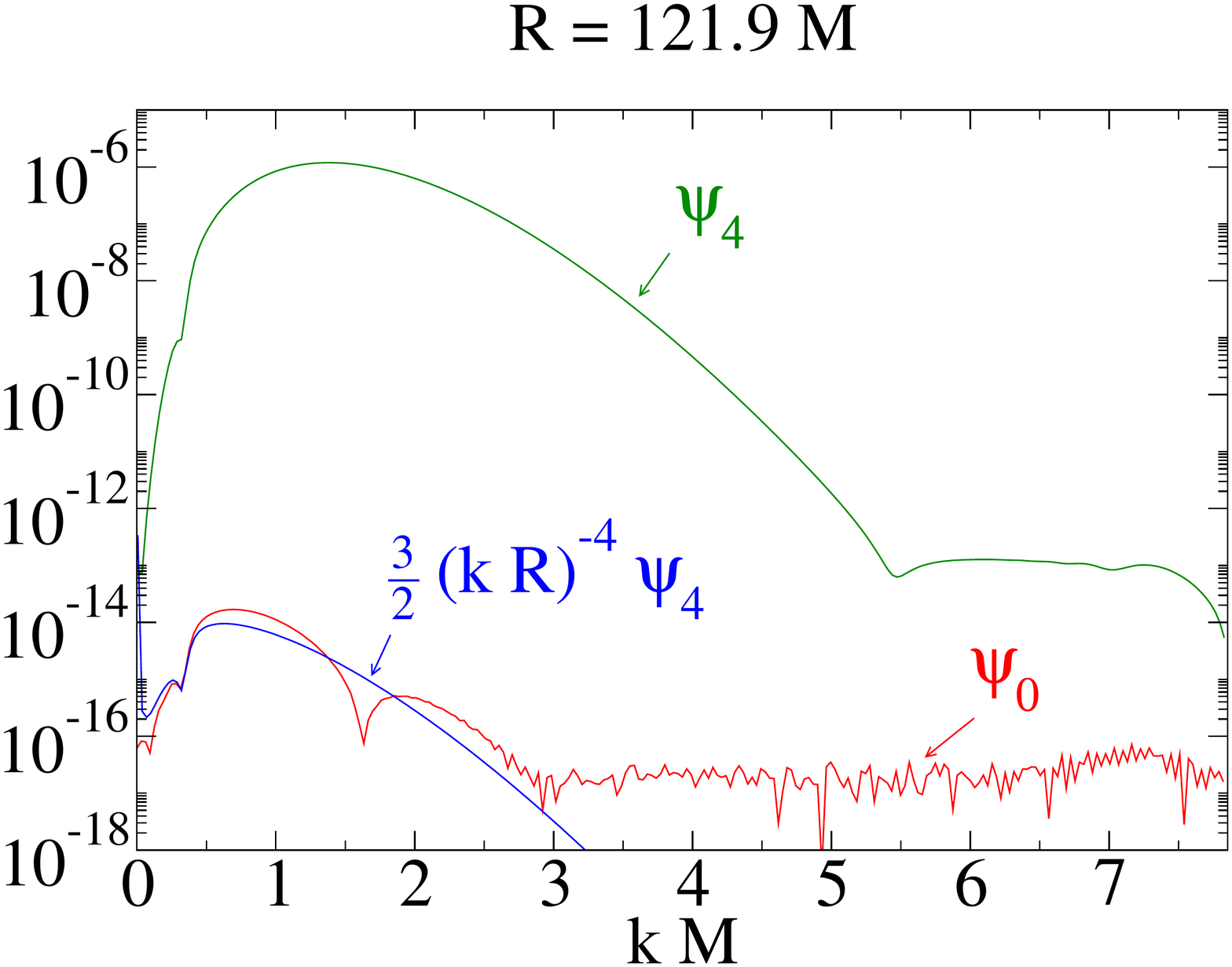}
  \caption{\label{f:PredictedPsi0} 
    Comparison of the time Fourier transform of
    the measured $\Psi_0(t)$ with $\frac{3}{2} (k R)^{-4} \Psi_4$,
    which is the predicted
    value using the reflection coefficient of
    \cite{Buchman2006}.
  }
\end{figure}

For a numerical solution using our new CPBCs, we evaluate the
Newman-Penrose scalars $\Psi_0(t)$ and $\Psi_4(t)$ on extraction
spheres located $1.9M$ inside the outer boundary.  
In figure \ref{f:PredictedPsi0} we plot the time Fourier transforms
of these quantities. We also plot $\frac{3}{2} (k R)^{-4} \Psi_4$, 
which by the above
argument should agree with $\Psi_0$ to leading order in $1/(k R)$.
Figure~\ref{f:PredictedPsi0} shows that the numerical agreement is 
reasonably good: roughly at the expected level of accuracy.
The overall dependence of the predicted
reflection coefficient $\rho$ on $k$ and $R$ is captured very well.
We surmise that the levelling off of our numerical $\Psi_0$ for $k
\gtrsim 3$ is due to numerical roundoff effects. (Note the magnitude
of $\Psi_0$ at those frequencies.)  For radii $R \gtrsim 200 M$,
$\Psi_0$ is at the roundoff level for all frequencies.

\section{Discussion}
\label{s:Discussion}

The purpose of this paper is to compare various methods of treating
the outer boundary of the computational domain.  We
evaluate the performance of several often-used boundary treatments in
numerical relativity by measuring the amount of spurious reflections
and constraint violations they generate.
To this end, we consider as a test problem an outgoing
gravitational wave superimposed on a Schwarzschild black hole spacetime.
First we compute this numerical solution on a reference domain,
large enough that the influence of the outer boundary can be neglected. 
Then we repeat the evolution on smaller domains using one of the 
boundary treatments, either imposing
local boundary conditions, compactifying the domain
using a radial coordinate map, or installing a sponge layer.
We use a first-order generalized harmonic formulation of the 
Einstein equations, although these boundary methods can be
applied to other formulations as well.  We believe our results are
fairly independent of the particular formulation used.

Our main conclusion is that our version of constraint-preserving
boundary conditions performs better than any of the alternate treatments
that we tested.  
Our boundary conditions include a limitation on the influx of spurious 
gravitational waves by freezing the Newman-Penrose scalar 
$\Psi_0$ at the boundary.  
We also introduce and test an improved boundary condition for the
gauge degrees of freedom.

For some of the simple boundary conditions, such as freezing or Sommerfeld
conditions, we find constraint violations that do not converge away
with increasing resolution.  The continuum limit does not
satisfy Einstein's equations in these cases.
Most of the alternate boundary conditions also generate considerable
reflections as measured by $\Delta \U$, the norm of the difference 
with respect to the reference solution.  In many cases, these 
reflections do not decrease significantly with increasing resolution.

The difference norm $\Delta \U$ that we use to measure boundary reflections
includes the entire spacetime metric, not just the physical degrees of
freedom.  It is important then to evaluate separately the effects of
the various boundary treatments on the physical degrees of freedom.
We use the extracted outgoing radiation as approximated by the
Newman-Penrose scalar $\Psi_4$ for this purpose.  Here our conclusions
are somewhat different.  Rather surprisingly, most of the boundary
methods we consider generate relatively small errors in $\Psi_4$.
This suggests that if gravitational waveforms are only needed to
an accuracy of, say, $1\%$ (which is comparable to the discrepancies
between recent binary black hole simulations \cite{Baker2007}) then
even the simple Sommerfeld conditions might be good enough. (For
those, we find relative errors $\sim 10^{-5}$.)  The largest
relative errors in $\Psi_4$ we find ($\sim 10^{-2}$)
occur with our implementation
of the spatial compactification method used by Pretorius
\cite{Pretorius2005a,Pretorius2005b,Pretorius2006}.  We attribute
these largely to the use of artificial dissipation.  Undesirable effects of
dissipation
might be somewhat less severe in binary black hole evolutions, which
have much larger wavelengths ($\lambda \sim 20-100M$) than ours
($\lambda \sim 4 M$).  Our tests suggest that the errors
in $\Psi_4$ can be made to decrease significantly with resolution only
by using more sophisticated constraint preserving and physical
boundary conditions.  The importance of using a physical boundary
condition on $\Psi_0$ is illustrated in particular 
by the difference between the
performance of our boundary conditions and those of Kreiss and
Winicour \cite{Kreiss2006}.

Some caveats regarding the interpretation of our results must be stated.
First, the ratio of the dominant wavelength to the radius of the
outer boundary is typically much larger for binary black hole 
evolutions (where $\lambda/R\gtrsim 0.5$) 
than for the simple test problem considered here  (where 
$\lambda/R\sim 0.1$). Boundary treatments generally
work better for smaller $\lambda/R$, i.e.~when the boundary is well out
in the wave zone.
Hence the results presented here are likely to be more accurate than
those from typical binary black hole simulations.
Second, we use spectral methods rather than finite-difference methods,
which are more commonly used in numerical relativity at this time.
This complicates the implementation of the kind of numerical
dissipation that is crucial for the
spatial compactification method to work. While we have attempted to
construct a filter that mimics the finite-difference dissipation as
closely as possible, a direct comparison is clearly impossible.
In finite-difference methods, the error introduced by the type of numerical
dissipation considered here is below the truncation error.  Hence
tests similar to ours but performed with a finite-difference method
would not be able to detect the effect of dissipation.

There are several directions in which the present work could be extended.
For large values of the outer boundary radius, we observe a
non-convergent power-law growth of the error in our test problem when
constraint-preserving boundary conditions are used; the origin of this 
growth should be investigated further.
It would be interesting to implement and test the hierarchy of physical 
boundary conditions that are perfectly absorbing for linearized gravity 
(including leading-order corrections due to the curvature and backscatter) 
found recently by Buchman and Sarbach \cite{Buchman2006,Buchman2007}.
Our boundary conditions could also be tested using
known exact solutions such as gauge waves, and comparisons could be made 
with the results
found in \cite{Babiuc2007}.

For completeness we also mention a number of additional outer boundary
approaches that were not addressed in this paper, but would also be
interesting future extensions of this research.  In \cite{Abrahams1988,
Abrahams1990}, boundary conditions for the full nonlinear Einstein
equations on a finite domain are obtained by matching to exact
solutions of the linearized field equations at the boundary.
Alternatively, the interior code could be matched to an `outer
module' that solves the linearized field equations numerically
\cite{Abrahams1998, Rupright1998, Rezzolla1999, Zink2006}.  Other
approaches involve matching the interior nonlinear Cauchy code to an
outer characteristic code (see \cite{Winicour2005} for a review) or
using hyperboloidal spacetime slices that can be compactified towards
null infinity (see \cite{Frauendiener2004} for a review).


\appendix

\section{Details on the numerical test problem}
\label{s:NumericalDetails}

\subsection{Initial data}
\label{s:InitialData}

The initial data used for our numerical tests are the same as in
\cite{Kidder2005}. The background solution is a Schwarzschild black
hole in Kerr-Schild coordinates,
\begin{equation}
  \rmd s^2 = -\rmd t^2 + \frac{2M}{r} (\rmd t + \rmd r)^2 + \rmd r^2 +
  r^2 \rmd \Omega^2.
\end{equation}
Throughout the paper, $M$ refers to the bare black hole mass of the
unperturbed background.
We superpose an odd-parity outgoing quadrupolar wave perturbation 
constructed using Teukolsky's method \cite{Teukolsky1982}. Its generating
function is taken to be a Gaussian $G(r) = A \exp [-(r - r_0)^2/w^2]$
with $A = 4 \times 10^{-3}$, $r_0 = 5M$, and $w = 1.5 M$.
The full non-linear initial value equations in the conformal thin
sandwich formulation are then solved to obtain initial data
that satisfy the constraints \cite{Pfeiffer2004}.  This yields initial
values for the spatial metric, extrinsic curvature, lapse function,
and shift vector.
We note that after the superposition, the resulting solution is still
nearly but not completely outgoing.

Our generalized harmonic formulation of Einstein's equations requires
initial data for the full spacetime metric and its first time
derivative. These can be computed from the 3+1 quantities obtained
above, provided we also choose initial values for the time derivatives
of the lapse function and shift vector.  These initial time
derivatives are freely specifiable and are equivalent to the initial
choice of the gauge source function $H_a$; we choose $\partial_t N=0$
and $\partial_t N^i=0$ at $t=0$.

\subsection{Numerical method}
\label{s:NumericalMethod}

We use a pseudospectral collocation method as described
for example in \cite{Kidder2005}.

The computational domain for the test problem considered here 
is taken to be a spherical shell extending
from $r = 1.9 M$ (just inside the horizon) out to some 
$r = R$. This domain is subdivided into spherical-shell subdomains
of extent $\Delta r = 10 M$.  
On each subdomain, the numerical solution is expanded in Chebyshev
polynomials in the radial direction and in spherical harmonics in the 
angular directions (where each Cartesian tensor component is expanded in 
the standard scalar spherical harmonics).
Typical resolutions are $N_r \in \{ 21, 31, 41, 51 \}$ coefficients 
per subdomain for the Chebyshev series and $l \leqslant L$ with 
$L \in \{ 8, 10, 12, 14 \}$ for the spherical harmonics.

We change the outer boundary radius $R$ by changing the number of
subdomains while keeping the width $\Delta r$ of each subdomain
fixed; this facilitates direct comparisons between runs with different
values of $R$. For example, the innermost four subdomains of the
reference solution (which has a total of 96 subdomains and $R=961.9M$)
are identical to the four subdomains used to compute the solution with
$R=41.9M$.

The evolution equations are integrated in time using a fourth-order 
Runge-Kutta scheme, with a Courant factor $\Delta t / \Delta x_\mathrm{min}$ 
of at most $2.25$, where $\Delta x_\mathrm{min}$ is the smallest distance
between two neighbouring collocation points. 
As described in \cite{Kidder2005}, the top four coefficients
in the \emph{tensor} spherical harmonic
expansion of each of our evolved quantities is set to
zero after each time step; this eliminates an instability associated with
the inconsistent mixing of tensor spherical harmonics in our approach.

We use two methods of numerically implementing boundary conditions;
the choice of method depends on the type of boundary conditions.
Boundary conditions that can be expressed as algebraic relations
involving the characteristic fields are implemented using a penalty
method (see \cite{Hesthaven2000} and references therein;
in the context of finite-difference methods see also \cite{Schnetter2006} 
and references therein).  
In particular, we use a penalty method to implement the Kreiss-Winicour
boundary conditions (cf. section \ref{s:KreissWinicour}) and to impose boundary
conditions at the internal boundaries between neighbouring subdomains.
Boundary conditions that are expressed in terms of the \emph{time
derivatives} of the characteristic fields are implemented using the
method of Bj{\o}rhus \cite{Bjorhus1995}, 
where the time derivatives of the
incoming characteristic fields are replaced 
at the boundary with the relevant boundary
condition.  All boundary conditions in this paper besides those mentioned
above are implemented using the Bj{\o}rhus method.

\subsection{Gauge source functions}

Our generalized harmonic formulation \cite{Lindblom2006} of Einstein's
equations allows for gauge source functions that
depend arbitrarily on the coordinates and the spacetime metric:
$H_a=H_a(t,x,\psi)$.  The generalized harmonic evolution equations
are equivalent to Einstein's equations only if the constraint (\ref{e:C1})
remains satisfied.

We choose the time derivatives of lapse and shift to be zero at the
beginning of the simulation; this determines the initial value of
$H_a$ via the constraint (\ref{e:C1}).  For the subsequent evolution,
we hold this $H_a$ fixed in time.

\subsection{Error quantities}
\label{s:ErrorQuantities}

We use two different measurements of the errors in our solutions, 
which we monitor during our numerical
evolutions.  First, given a numerical solution 
$(\psi_{ab}, \Pi_{ab}, \Phi_{iab})$, the difference between that solution 
and the reference solution 
$(\psi_{ab}^{\mathrm{(ref)}}, \Pi_{ab}^{\mathrm{(ref)}},
\Phi_{iab}^{\mathrm{(ref)}})$ is computed with the following
norm at each point in space,
\begin{eqnarray}
  \label{e:DeltaU}
  \Delta \U &\equiv& 
\left[ \delta^{ab} \delta^{cd} ( 
     M^{-2} \Delta \psi_{ac} \Delta \psi_{bd} 
     + \Delta \Pi_{ac} \Delta \Pi_{bd} 
\right. \nonumber \\ && \left. 
+ g^{ij} \Delta \Phi_{iac} \Delta \Phi_{jbd} ) \right]^{1/2} ,
\end{eqnarray}
where $\Delta \psi_{ab}$ means $\psi_{ab}-\psi_{ab}^{\mathrm{(ref)}}$, and
similarly for $\Delta \Pi_{ab}$ and $\Delta \Phi_{iab}$.
Second, we define a quantity $\C$ that measures the violations
in all of the constraints of our system,
\begin{eqnarray}
  \C &\equiv& \left[ \delta^{ab} (
 \F_a \F_b + g^{ij} (\C_{ia} \C_{jb} + g^{kl} \delta^{cd} \C_{ikac} \C_{jlbd}) 
    \right. \nonumber\\ && \left.     
    + M^{-2} (\C_a \C_b + g^{ij} \delta^{cd} \C_{iac} \C_{jbd}) 
    ) \right]^{1/2} ,
\end{eqnarray}
where $\F_a$ and $\C_{ia}$ are first derivatives of $\C_a$ defined in
\cite{Lindblom2006}.
To compute global error measures, a spatial norm $||\cdot||$,
either the $L^\infty$ norm or the $L^2$ norm, is applied
separately to $\Delta \U$ and $\C$.

The question often arises as to the significance of particular values
of $||\Delta \U||$ and $||\C||$. For example, is a
simulation with $||\C||=10^{-2}$ good to one percent accuracy? To make
it easier to answer such questions, we normalize both $||\Delta \U||$
and $||\C||$ as follows, and we always plot normalized quantities.

We divide $||\Delta \U||$ by a normalization factor $||\Delta \U_0||$,
defined as the difference between a given solution at $t=0$ and the
unperturbed Schwarzschild background; i.e., the quantity $||\Delta \U_0||$
is computed from \eref{e:DeltaU} using the unperturbed Schwarzschild
solution instead of the reference solution.  Since $||\Delta \U_0||$
is evaluated at $t=0$, it depends only on the initial data used in the
simulation, and is a measure of the amplitude of the superposed
gravitational wave perturbation. 
For the initial data used here, $||\Delta \U_0||_\infty =
6 \times 10^{-3}$ and $||\Delta \U_0||_2 = 1.4 \times 10^{-4}$.  The
quantity $||\Delta \U||/||\Delta \U_0||$ is more easily interpreted
than $||\Delta \U||$; for example, $||\Delta \U||/||\Delta \U_0||$ is unity
when the difference from the reference solution is of the same size 
as the initial perturbation.

Similarly, the constraint energy norm $||\C||$ is divided by the norm of the
first derivatives $||\partial \U||$ (at the respective time),
\begin{eqnarray}
  \partial \U &\equiv& \left[ g^{ij} \delta^{ab} \delta^{cd} ( 
     M^{-2} \partial_i \psi_{ac} \partial_l \psi_{bd} 
     + \partial_i \Pi_{ac} \partial_j \Pi_{bd} 
     \right. \nonumber\\ && \left.   
     + g^{kl} \partial_i \Phi_{kac} \partial_j \Phi_{lbd} ) \right]^{1/2} .
\end{eqnarray}
The constraints for our system are linear combinations of the
first derivatives of the fields, hence $||\C||/ ||\partial \U|| \sim 1$ 
corresponds to a complete violation of the constraints.

\subsection{Wave extraction}
\label{s:WaveExtraction}

For evaluating gravitational waveforms, we compute the Newman-Penrose scalars
\begin{equation}
  \Psi_0 = -C_{abcd} l^a m^b l^c m^d , \qquad
  \Psi_4 = -C_{abcd} k^a \bar m^b k^c \bar m^d , 
\end{equation}
where $C_{abcd}$ is the Weyl tensor, $l^a$ and $k^a$ are outgoing and ingoing
null vectors normalized according to $l^a k_a=-1$,
$m^a$ is a complex unit null spatial vector orthogonal to
$l^a$ and $k^a$, and $\bar m^a$ is the complex conjugate of $m^a$.
For perturbations of flat spacetime, there is a standard choice
for the vectors $l^a$, $k^a$, and $m^a$. 
In general curved spacetimes, however, no such prescription for the
tetrad exists that would produce coordinate-independent quantities
$\Psi_0$ and $\Psi_4$ at finite radius.
We choose the null vectors according to
\begin{equation}
       l^a = \textstyle \frac{1}{\sqrt{2}}\left(t^a + n^a\right), \qquad 
       k^a = \textstyle \frac{1}{\sqrt{2}}\left(t^a - n^a\right),
\end{equation}
where $t^a$ is the future-pointing unit timelike normal to the $t =
\mathrm{const.}$ slices and $n^a$ is the unit spacelike normal to the
extraction sphere. Finally, we choose
\begin{equation}
m^a = \frac{1}{\sqrt{2} r}
            \left(\frac{\partial}{\partial \theta} 
            +    i\frac{1}{\sin\theta}\frac{\partial}{\partial \phi}\right)^a,
\end{equation}
where $(r,\theta,\phi)$ are spherical coordinates on the 
$r = R_\mathrm{ex} = \mathrm{const.}$ extraction sphere.  
Note that our choice of $m^a$ is not exactly null nor of unit magnitude 
at finite extraction radius.
However, the tetrad is orthonormal for the unperturbed Schwarzschild
solution, so that the errors introduced into $\Psi_0$ and $\Psi_4$
because of the lack of tetrad orthonormality will be second-order
small in perturbation theory. 

The quantity $\Psi_4$ corresponds to outgoing radiation in the limit of 
$r \rightarrow \infty$, $t - r = \mathrm{const.}$, i.e.~as future 
null infinity is approached.  Similarly $\Psi_0$ corresponds 
to ingoing radiation as past null infinity is approached.
At finite extraction radius, $\Psi_4$ and $\Psi_0$ will disagree with
the waveforms observed at infinity by terms of the order
$\Or(R_\mathrm{ex})^{-1}$.

We decompose the quantities $\Psi_4$ and $\Psi_0$ in terms of
spin-weighted spherical harmonics of spin-weight $-2$ on the extraction
surface. 
Since our perturbation is an odd-parity quadrupole wave, the imaginary
part of the $(l = 2, \, m = 0)$ spherical harmonic is by far the dominant 
contribution to $\Psi_4$, and we only display that mode in our plots.
We normalize the curves in our graphs by the maximum (in time) value
of $|\Psi_4|$ at the extraction radius $R_\mathrm{ex}$, which for 
$R_\mathrm{ex} = 40 M$ is $\max |\Psi_4| = 6 \times 10^{-4}$.


\section{Details of the alternate approaches}
\label{s:AltDetails}

In this appendix, we provide some more details on the alternate
boundary treatments discussed in section \ref{s:AltApproach}: spatial
compactification and sponge layers. 

\subsection{Spatial compactification}
\label{s:CompactDetails}

\begin{figure}
  \plot{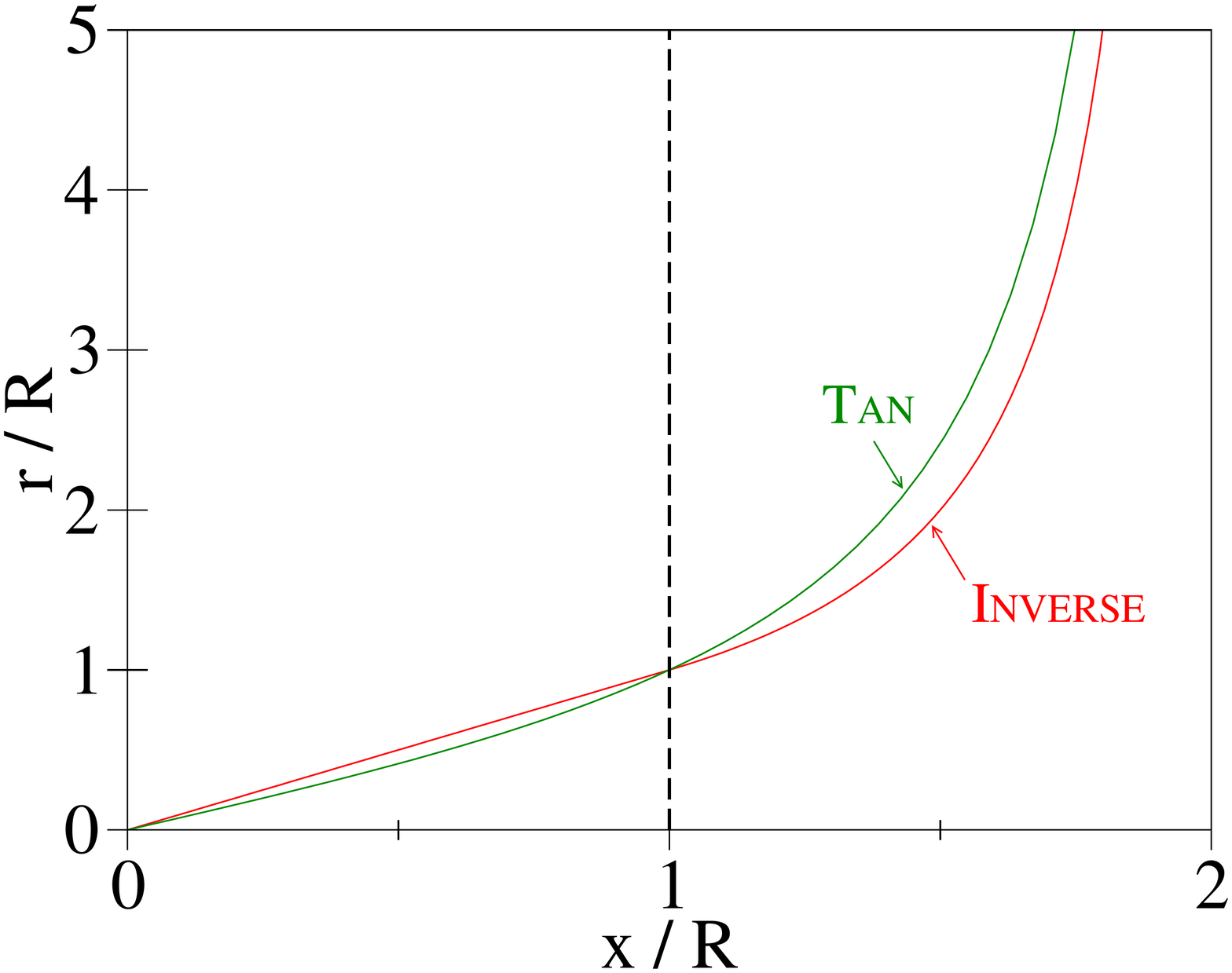}
  \plot{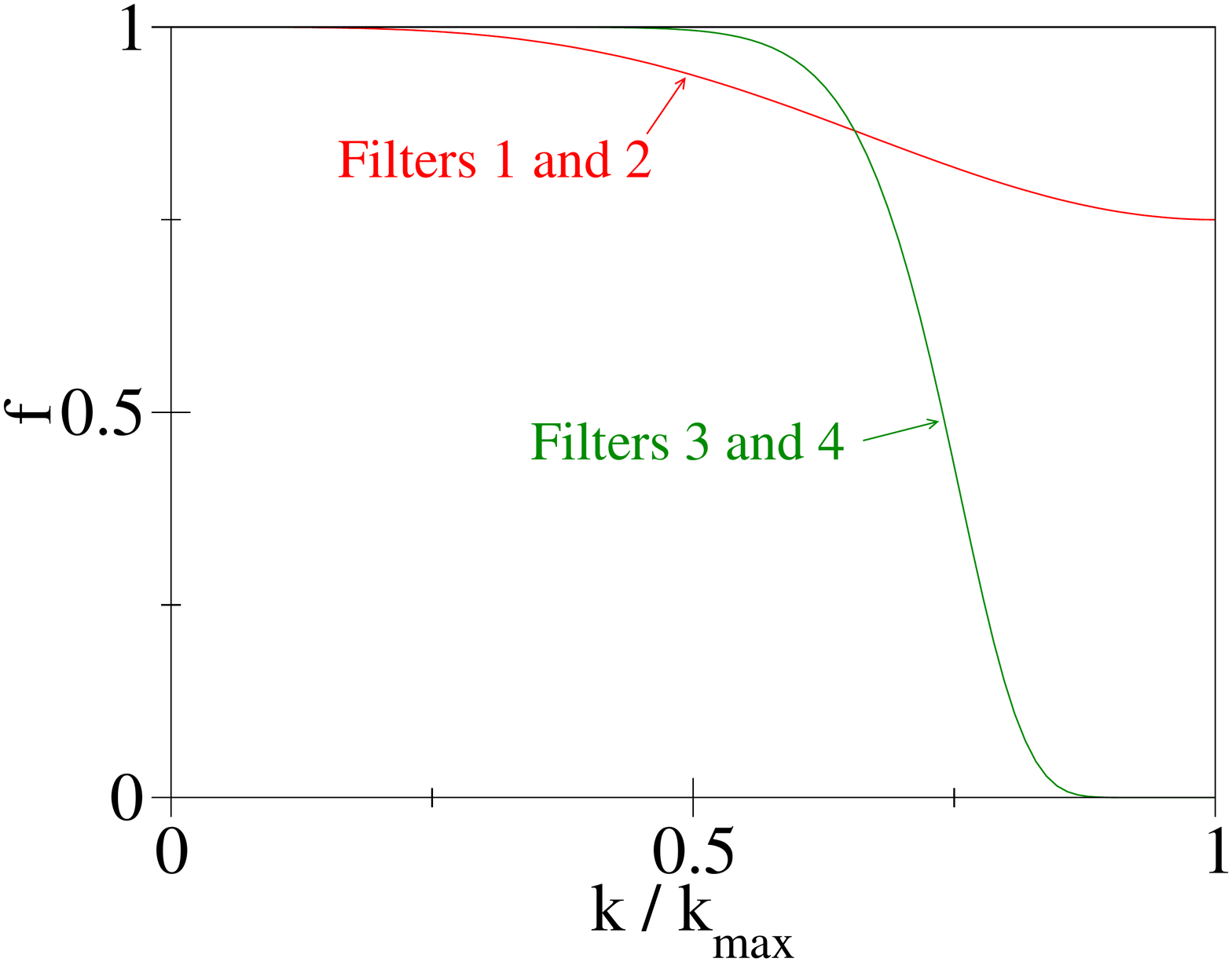}
  \caption{\label{f:MapsAndFilters} 
    Compactification mappings (left) and filter functions (right).
    The dashed line indicates the boundary of the region in where the
    compactification mapping is (essentially) linear.}
\end{figure}

We implement spatial compactification by introducing a radial coordinate
transformation $x \rightarrow r(x)$
that maps a compact ball on the computational grid with 
$x \in [0, x_\mathrm{max}]$ to the full unbounded physical slice with
$r \in [0, \infty]$.  We consider two such mappings.
The \textsc{\small Tan} mapping is similar to the one used by Pretorius 
\cite{Pretorius2005a,Pretorius2005b,Pretorius2006} and is given by
\begin{equation}
  r_\mathrm{\textsc{\small Tan}}(x) 
                       = R \tan \left( \frac{\pi x}{4 R} \right) , \qquad 
  0 \leqslant x < 2 R .
\end{equation}
The scale $R$ determines the range in physical radius $r$ across which
the map is essentially linear (see figure \ref{f:MapsAndFilters}).
When comparing compactification with other boundary treatments, we
compare quantities only in the region $r<R$.
(The scale $R$ is equal to unity in the work of Pretorius. He uses 
mesh refinement to obtain the appropriate resolution close to
the origin, while we fix the resolution and choose the scale $R$
appropriately.)
We also tested an \textsc{\small Inverse} map defined by
\begin{equation}
  r_\mathrm{\textsc{\small Inverse}}(x) = \cases{
    x , & $0 \leqslant x \leqslant R$ , \\ 
    \frac{R^2}{2R - x} , & $R < x < 2R$,
  }
\end{equation}
see figure \ref{f:MapsAndFilters}.  This map is only $C^1$ at $x=R$,
but we maintain spectral accuracy in our tests by placing this
surface at the boundary between spectral subdomains. 

Dissipation is needed to remove the short wavelength components of the waves
as they travel outward on the compactified computational grid and become
unresolved.
We apply this dissipation only in the radial direction,
but everywhere in the computational domain.
In spectral methods, dissipation can be conveniently implemented in the
form of a spectral \emph{filter}.
This filter is applied by multiplying each spectral expansion coefficient 
of index $k$ by a function $f(k)$.
(See \ref{s:NumericalMethod} for details on the pseudospectral method 
we use.) 
Higher values of $k$ correspond to shorter wavelengths 
in the numerical approximation; let $k_\mathrm{max}$ be the highest 
index used in the spectral expansion.
The first filter function we consider is the 
closest analogue in the context of our spectral methods 
to Kreiss-Oliger \cite{KreissOliger1973} dissipation,
\begin{equation}
  \label{e:KreissOligerFilter}
  f_\mathrm{\textsc{\small Kreiss-Oliger}}(k) = 
  1 - \epsilon \, \sin^4 \left( \frac{\pi k}{2 k_\mathrm{max}} \right), 
  \qquad 0 \leqslant \epsilon \leqslant 1 .
\end{equation}
Typical values of the parameter $\epsilon$ used by Pretorius are
$\epsilon \in [0.2, \, 0.5]$; we use $\epsilon = 0.25$.

This filter was derived via a comparison with finite-difference
methods as follows. 
In the finite-difference approach, a numerical solution $u$ is
represented on a set of equidistant grid points $x_j$.  
(It suffices to consider the one-dimensional case here.)
Some form of numerical dissipation is usually required for the
finite-difference method to be stable. The one that is most often used
for second-order accurate methods is fourth-order Kreiss-Oliger 
dissipation \cite{KreissOliger1973}. One possible implementation of
this, used e.g.~by Pretorius, amounts to replacing
\begin{equation}
  u \rightarrow F[u] \equiv \left( 1 - \frac{\epsilon}{16} h^4 D^4 \right) u 
\end{equation}
at each time step, 
where $h$ is the grid spacing and $D^4$ is the second-order accurate 
centred finite difference operator approximating the fourth derivative,
\begin{equation}
  D^4 u_i = h^{-4} (u_{j-2} - 4 u_{j-1} + 6 u_j - 4 u_{j+1} + u_{j+2}).
\end{equation}
Taking $u$ to be a Fourier mode $u^{(k)}_j = \exp (\rmi k x_j)$, 
it follows that the mode is damped by a frequency-dependent factor,
\begin{equation}
  \label{e:KreissOligerFilter1}
  u^{(k)} \rightarrow F[u^{(k)}] \equiv \left[ 1 - \epsilon \, \sin^4 
    \left( \frac{\pi k}{2 k_\mathrm{max}} \right) \right] u^{(k)} ,   
\end{equation}
where $k_\mathrm{max} = \pi / (2h)$ is the Nyquist frequency.
Thus we obtain the filter function \eref{e:KreissOligerFilter}.
Strictly speaking, the above analysis only applies to Fourier expansions 
and not to the Chebyshev expansions we use. Nevertheless, we apply the filter
in the form \eref{e:KreissOligerFilter} to our Chebyshev expansion
coefficients.  
Note that in \eref{e:KreissOligerFilter1}, each spectral coefficient
$u^{(k)}$ is filtered separately; this is not true for the analogous
calculation for a Chebyshev expansion.

We also use a different filter function, which we call the
\textsc{\small Exponential} filter,
that is often used in spectral methods (see \cite{Gottlieb2001} and
references therein),
\begin{equation}
  f_\mathrm{\textsc{\small Exponential}}(k) = 
  \exp \left[ - \left( \frac{k}{\sigma k_\mathrm{max}} \right)^p
  \right] .
\end{equation}
Typical values of the parameters are $\sigma = 0.76$ and $p = 13$.
This choice of parameters gives less dissipation at small values
of $k$ than the Kreiss-Oliger filter, and also ensures 
that $f(k_\mathrm{max}) \approx
10^{-16}$ is at the level of the numerical roundoff error.

There are various ways the filters can be applied in a
numerical evolution. We have experimented with two different methods.
In the first method, the filter is applied to the right side of the
equations,
i.e.~the evolution equations $\partial_t u = S$ 
are modified according to $\partial_t u = F[S]$, where
$F[S]$ is the filtered right side.  
In the second method, the filter is instead applied to the solution itself, 
i.e.~after each substep of the time integrator (cf. \ref{s:NumericalMethod}), 
the numerical solution $u$ is replaced with its filtered version $F[u]$. 
This second method
is closest to how the Kreiss-Oliger filter is applied by Pretorius.

For our numerical tests, we have used four different combinations of
the various options described above. They are summarized in 
table \ref{t:Filtering}.

\begin{table}
  \centering
  \begin{tabular}{|l|l|l|l|}
    \hline
    No. & Type & Parameters & Applied to \\
    \hline
    1 & \textsc{\small Kreiss-Oliger} & $\epsilon = 0.25$ & right side \\
    2 & \textsc{\small Kreiss-Oliger} & $\epsilon = 0.25$ & solution \\
    3 & \textsc{\small Exponential} & $\sigma = 0.76$, $p = 13$ & right side \\
    4 & \textsc{\small Exponential} & $\sigma = 0.76$, $p = 13$ & solution \\
    \hline
  \end{tabular}
  \caption{\label{t:Filtering} 
    Details of the filtering methods
  }
\end{table}

\subsection{Sponge layers}
\label{s:SpongeDetails}

For sponge layers we must specify a sponge profile function $\gamma(r)$,
as defined in \eref{e:spongemodification}. We choose $\gamma(r)$
to be nonzero only outside some sponge-free region of radius $R$,
and when comparing sponge layers with other boundary treatments,
we compare quantities only in the sponge-free region $r<R$.

\begin{figure}
  \centering
  \plot{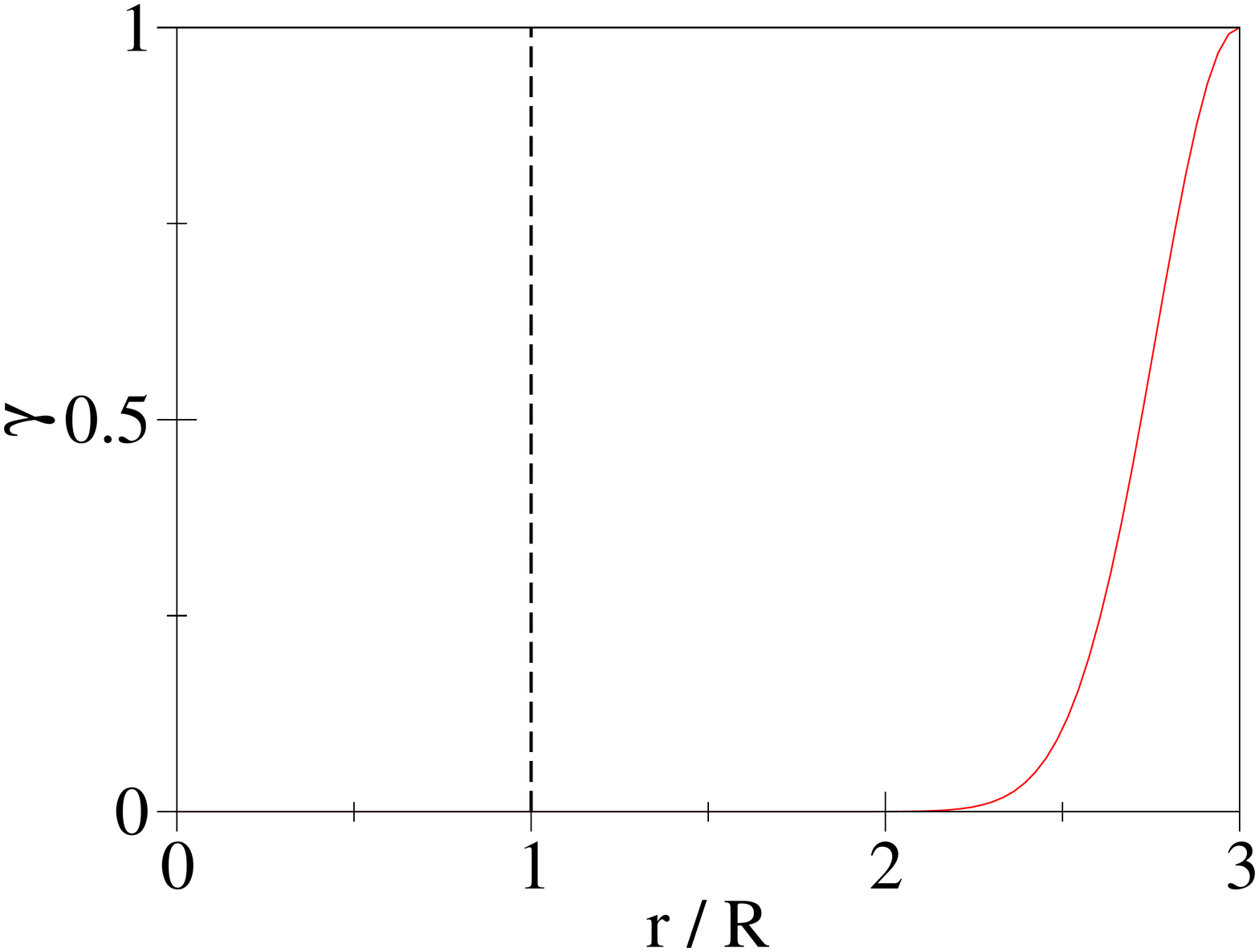}
  \caption{\label{f:SpongeProfile} 
    The sponge profile function $\gamma(r)$. The dashed line indicates
    the boundary of the region where $\gamma$ is below the numerical 
    roundoff error.}
\end{figure}

The sponge profile function $\gamma(r)$ we use is a Gaussian
centred at the outer boundary, which we choose to place at $r = 3R$, 
\begin{equation}
  \gamma(r) = \gamma_0 \exp \left[ - \left( \frac{r - 3 R}{\sigma} \right)^2
  \right] .
\end{equation}
The amplitude of the Gaussian is taken to be $\gamma_0 = 1$.
The width $\sigma$ is chosen so that $\gamma(r) \leqslant 10^{-16}$
(the numerical roundoff error) for $r \leqslant R$, which requires
$\sigma \lesssim R/3$.  
In our numerical example, we take $R = 41.9 M$ and $\sigma = 13.3 M$.
Hence $\sigma$ is considerably larger than the wavelength 
$\lambda \approx 4 M$ of the gravitational wave, which is required in order 
to avoid reflections from the sponge layer (cf. section 17.2.3 of 
\cite{Boyd2001}).  
Figure \ref{f:SpongeProfile} shows a plot of this sponge profile.

\ack We thank Luisa Buchman, Jan Hesthaven, Larry Kidder, Harald
Pfeiffer, Olivier Sarbach, and Jeff Winicour for helpful discussions 
concerning this work.  
The numerical simulations presented here were performed using the 
Spectral Einstein Code (SpEC) developed at Caltech and Cornell
primarily by Larry Kidder, Mark Scheel and Harald Pfeiffer.  This work
was supported in part by grants from the Sherman Fairchild Foundation,
and from the Brinson Foundation; by NSF grants PHY-0099568,
PHY-0244906, PHY-0601459, DMS-0553302 and NASA grants NAG5-12834,
NNG05GG52G.


\section*{References}

\bibliographystyle{iopart-num}
\bibliography{References}

\end{document}